\documentclass[num-refs]{ametsocV6.1}

\newcommand\BibTeX{{\rmfamily B\kern-.05em \textsc{i\kern-.025em b}\kern-.08em
		T\kern-.1667em\lower.7ex\hbox{E}\kern-.125emX}}

\usepackage{moreverb}

\usepackage[utf8]{inputenc}
\usepackage[T1]{fontenc}

\usepackage{amsmath}
\usepackage[hidelinks]{hyperref}
\usepackage{amssymb}
\usepackage{esint} 

\usepackage{graphicx}
\usepackage{placeins}
\usepackage{tikz,pgfplots}
\usetikzlibrary{positioning}

\usepackage[caption=false]{subfig}
\usepackage{float}

\usepackage{siunitx}
\usepackage{color}
\usepackage{eurosym}
\usepackage[colorinlistoftodos]{todonotes}

\title{Multi-stability of Atlantic and Pacific overturning: The role of Freshwater Forcing Asymmetries and the Hydrological Cycle}
\clearpage
\authors{Elian Vanderborght\aff{a}\correspondingauthor{Elian Vanderborght, e.y.p.vanderborght@uu.nl}, Oliver Mehling\aff{a}
	Henk A. Dijkstra \aff{a,b} }

\affiliation{\aff{a}Institute for Marine and Atmospheric research Utrecht, Department of Physics, 
	Utrecht University, Utrecht, the Netherlands\\
	\aff{b} Centre for Complex Systems Studies,  Utrecht University, Utrecht, the Netherlands. }
\abstract{A defining feature of the present-day global overturning circulation (GOC) is the absence of deep water formation in the Pacific, in contrast to the Atlantic. This asymmetry, associated with higher surface salinities in the North Atlantic, is reflected in the Atlantic Meridional Overturning Circulation (AMOC) and the lack of a Pacific overturning (PMOC). A commonly cited explanation is the asymmetry in surface freshwater fluxes, with the Pacific receiving more freshwater per unit area than the Atlantic. Here, we develop a two-basin conceptual ocean model, consisting of a wide and a narrow basin. The model admits three states: sinking confined to the narrow basin, sinking confined to the wide basin, and sinking in both basins. We analyze the (co-)existence of these states as a function of freshwater asymmetry and hydrological cycle strength, defined as the longitudinally symmetric freshwater flux. For a weak hydrological cycle, representative of warm Pliocene-like climate conditions, sinking occurs in both basins, with symmetry breaking only when one basin is sufficiently more evaporative. For intermediate conditions, representative of the present-day climate, the basin with slightly stronger evaporation tends to host sinking, with a stronger preference in the narrow basin. For a strong hydrological cycle, single-basin sinking states are preferred, although a large interbasin freshwater asymmetry is required to uniquely localize sinking. These results provide insight into GOC sinking configurations under past, present, and potential future climates, and show good agreement with a three-dimensional global circulation model.}
\begin{document}		
	\maketitle
	\statement{A defining question in physical oceanography is why deep water formation occurs in the Atlantic (AMOC) but not in the Pacific (PMOC). A common explanation is that the Pacific receives more freshwater, suppressing sinking, but this remains debated due to the many other proposed mechanisms. We show that this freshwater asymmetry is essential under present-day conditions, explaining why the Atlantic hosts the AMOC rather than the PMOC. Its influence, however, depends strongly on the intensity of the hydrological cycle. For weak cycles, sinking occurs in both basins, whereas for strong cycles it is confined to a single basin. In both cases, the freshwater asymmetry exerts only limited control on the localization of sinking. These results suggest that the role of freshwater fluxes in setting overturning patterns varies across climate states, with important implications for past and future ocean circulation.}
	\newpage
	\section{Introduction}
	The global overturning circulation (GOC) plays a central role in redistributing heat, carbon, and salt within the global ocean. Below the Ekman layer, and extending to depths of approximately 3,000 m, it includes the so-called mid-depth cell \citep{cessi2019global}. This cell is characterized by deep water formation in the high-latitude Atlantic, where relatively salty surface waters lose buoyancy and are transformed into North Atlantic Deep Water (NADW). NADW forms the lower limb of the Atlantic Meridional Overturning Circulation (AMOC) and flows southward at depths between 1,000 and 3,000 m \citep{lumpkin2007global}.
	
	In contrast, the present-day Pacific lacks an equivalent formation of North Pacific Deep Water (NPDW) and a corresponding Pacific Meridional Overturning Circulation (PMOC). The origin of this interbasin asymmetry remains a longstanding question in dynamical oceanography (see \citealp{ferreira2018atlantic} for a review). It has major implications for global meridional heat transport \citep{ferrari2011processes} and, consequently, for regional climate (e.g. \citealp{van2025physics}). The absence of NPDW formation is commonly attributed to the low sea-surface salinity and pronounced halocline in the North Pacific, which together create a strongly stratified upper ocean that suppresses deep convection \citep{warren1983there,haug1999onset}. The interbasin contrast in northern sea-surface salinity has, in turn, been linked to asymmetries in atmospheric surface forcing, basin geometry, and the oceanic circulation itself.
	
	Atmospheric surface-forcing asymmetries arise in part because the North Pacific receives more freshwater per unit basin width than the North Atlantic \citep{emile2003warren,seidov2005there,craig2017contrast}. This freshwater contrast results from moisture transport by both stationary (e.g. \cite{sinha2012mountain}) and transient (e.g. \cite{ferreira2010localization}) atmospheric eddies \citep{wills2016stationary}. Additional contributions arise from anomalous moisture convergence in the tropical Pacific, associated with eastward transport across Panama \citep{zaucker1994atmospheric}, and westward transport across the western Pacific boundary linked to the monsoonal circulation \citep{emile2003warren}. Surface wind variability further amplifies this asymmetry: stronger variability over the Atlantic shortens mixing timescales, promoting higher surface salinity relative to the Pacific \citep{ferreira2018atlantic}.
	
	The role of interbasin freshwater flux asymmetry on the GOC has been investigated in numerous modeling studies. For example, \cite{menviel2012removing} showed that compensating for the excess freshwater flux over the Pacific leads to the development of an active PMOC and a weakening of the AMOC. Moreover, studies examining the influence of orography on the GOC find that reduced orography weakens or even collapses the AMOC, while strengthening the PMOC. This response similarly arises from changes in the surface freshwater flux asymmetry. In particular, lowering the Tibetan Plateau decreases the freshwater flux over the Pacific relative to the Atlantic \citep{su2018difference,yang2020investigating,yang2024north}, while lowering the Rocky Mountains enhances moisture transport from the Pacific to the Atlantic \citep{schmittner2011effects,maffre2018influence}. Both mechanisms reduce, and may even reverse, the interbasin freshwater flux asymmetry, yielding a modified GOC configuration.
	
	Asymmetries in ocean basin geometry may further favor sinking in the Atlantic. In particular, basin width plays an important role: narrower basins are more conducive to deep-water formation because northward western boundary salt transport is less effectively compensated by the southward western boundary flow of the subpolar gyre \citep{de2008atlantic,jones2017size}. Other geometric features that facilitate sinking in the Atlantic include its more northward extension, which enhances access to cold high-latitude surface waters \citep{de2008atlantic,huisman2012does}; interbasin exchanges through seaways such as the Mediterranean and the Bering Strait \citep{reid1979contribution,hu2012pacific}; and the shorter southward extent of the eastern continent of the Atlantic basin, which, equatorward of the zero Ekman pumping line, yields a saline westward transport from the Pacific to the Atlantic \citep{nilsson2013ocean,cessi2019global}.
	
	However, asymmetries in sinking location may also emerge spontaneously through advective feedbacks \citep{marotzke1991multiple}. In particular, the elevated salinity of the North Atlantic can result from the circulation itself, which imports high-salinity subtropical waters into the basin. In this framework, the presence of sinking in one basin rather than the other depends on the initial conditions \citep{huisman2009robustness, nilsson2013ocean}. 
	
	Proxy records suggest that an active PMOC existed during the warm Pliocene (5.3–2.6~Ma ago) \citep{kwiek1999pacific,burls2017active,ford2022sustained}—a period with a continental configuration similar to today’s, apart from differences in ocean gateways. In particular, \cite{burls2017active} argued that overturning symmetry, i.e., the coexistence of an active AMOC and PMOC, prevailed during this period due to a weakened hydrological cycle. Additional support for PMOC activation during the last deglaciation (17.5–15~kyr ago) has also been found \citep{hu2012pacific}, which was similarly attributed to changes in the hydrological cycle \citep{okazaki2010deepwater}. Moreover, recent studies indicate that a PMOC may spontaneously develop under greenhouse gas forcing \citep{curtis2024spontaneous}, which is associated with a stronger weakening of the AMOC under such forcing \citep{baker2025continued}.
	
	These findings suggest that the GOC may adopt different sinking configurations under past and future climate conditions. To understand which configuration is preferred, we must determine how interbasin asymmetries bias the system toward a particular sinking state and how this sensitivity depends on the background climate. 
	
	Here, we focus on asymmetries in surface freshwater forcing and their role in breaking overturning symmetry and biasing the system toward sinking in a particular basin. However, in the presence of the salt-advection feedback, multiple equilibria may arise. The realized overturning state may then depend on the initial conditions, complicating efforts to isolate how interbasin freshwater fluxes bias the system toward a given configuration.
	
	In this study, we revisit this problem using a conceptual model building on earlier work by \cite{gnanadesikan1999simple,jones2016interbasin,gnanadesikan2024tipping}. The model represents two basins—one narrow and one wide—connected by a zonally periodic re-entrant channel. Within this idealized framework, we isolate the role of surface freshwater-flux asymmetry in determining a preferred sinking configuration and examine how this preference depends on key parameters of the background climate state, focusing on the magnitude of the longitudinally symmetric freshwater forcing, here interpreted as a measure of hydrological cycle intensity. Using continuation techniques, we systematically trace all equilibrium branches, thereby avoiding the interpretational challenges associated with the salt-advection feedback.
	
	The paper is organized as follows. In Section~\ref{S:M}, we describe the conceptual model, its underlying assumptions, and key results from the model. These results are compared with the output of GCM simulations in Section~\ref{S:CCM}. Finally, in Section~\ref{S:S&D}, we summarize our conclusions, discuss their broader implications, and place our findings in the context of existing literature.
	
	\section{Conceptual Model}\label{S:M}
\begin{figure}
		\captionsetup{justification=centering}
		\centering
		\includegraphics[width=0.5\linewidth]{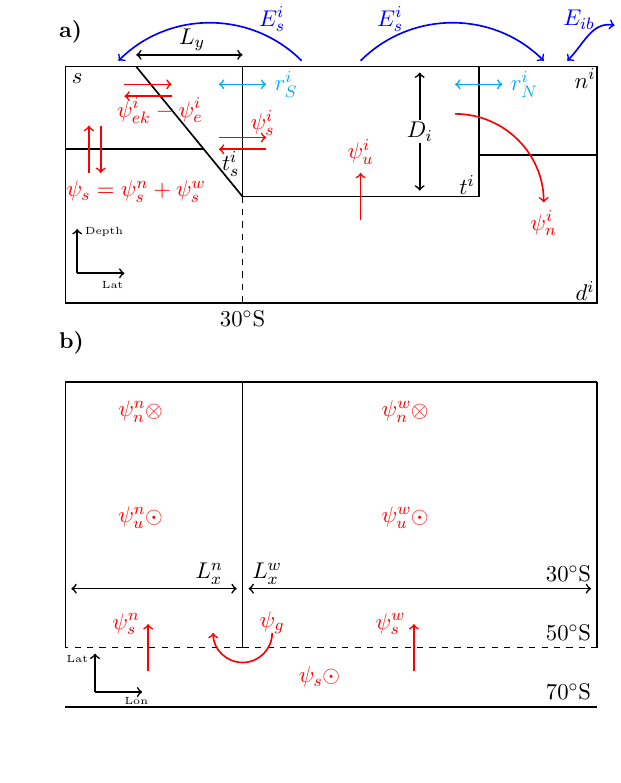}
		\caption{Schematic of a two-basin box model (CTBM): (a) latitude–depth cross-section, and (b) longitude–latitude cross-section along the upper interface of the deep box. Each basin consists of four distinct boxes and a shared southern box, $s$. Volume, freshwater, and gyre fluxes between boxes are indicated in red, blue, and cyan, respectively. In panel (b), $\otimes$ and $\odot$ denote downward and upward fluxes through the interface, and all fluxes are shown as positive. A detailed description of the volume, freshwater, and gyre fluxes is provided in the main text.}
		\label{fig:Model_Schematic}
	\end{figure}

	The model of \cite{gnanadesikan1999simple} represents the ocean as a two-layer system, in which the interface—interpreted here as the pycnocline depth—separates intermediate and deep waters. Thickness is exchanged between the layers through northern sinking, diapycnal upwelling, and Ekman and eddy fluxes. This provides a crude representation of the upper branch of the Residual Overturning Circulation (ROC) within a single basin. \cite{jones2016interbasin} extended this framework to two basins, while restricting Northern Hemisphere sinking to one basin. The pycnocline volume budgets were modified to include a geostrophic exchange flux at the southern boundary of the interface connecting the semi-enclosed basins. \cite{gnanadesikan2024tipping} further generalized the model by allowing high-latitude sinking in both basins. The model presented here is a simplified and slightly modified version of this framework. A schematic is shown in Fig.~\ref{fig:Model_Schematic}.
	
	The model features two basins (Fig.~\ref{fig:Model_Schematic}b): a narrow basin representing an Atlantic analogue and a wide basin representing a Pacific analogue. Following \cite{cimatoribus2014meridional}, each basin is divided into four boxes (Fig.~\ref{fig:Model_Schematic}a): a variable-volume pycnocline box, itself composed of a $t_s^i$ box and a $t^i$ box; a fixed-volume northern box ($n^i$); and a variable-volume deep box ($d^i$), which separates the deep ocean from the pycnocline. Here, the superscript $i \in \{n,w\}$ denotes the basin, with $n$ and $w$ referring to the narrow and wide basins, respectively. Both basins are connected to a zonally periodic southern box ($s$), an analogue of the Southern Ocean, characterized by longitudinally invariant dynamics. The model therefore comprises nine boxes in total and can be viewed as a global, two-basin, extension of the model of \cite{cimatoribus2014meridional}, building on \cite{jones2016interbasin}, and is hereafter referred to as the Conceptual Two Basin Model (CTBM).
	
	\subsection{Pycnocline and Transport Equations}\label{S:MM}
	The pycnocline volume budget of the basins are given by:
	\begin{subequations}
		\begin{align}
			\left(\frac{L_y L_x}{2}+A_n\right)\frac{\partial D_n}{\partial t} &= \psi_s^n + \psi_u^n - \psi_n^n + \psi_g, \label{Dn} \\
			\left(\frac{L_y L_x^w}{2}+A_w\right)\frac{\partial D_w}{\partial t} &= \psi_s^w + \psi_u^w - \psi_n^w - \psi_g. \label{Dw}
		\end{align}
	\end{subequations}
	Here, $D_n$ and $D_w$ denote the pycnocline depths of the narrow and wide basins, respectively. $L_y$ is the meridional extent of the $t_s^n$ and $t_s^w$ boxes, while $L_x^n$ and $L_x^w$ represent the widths of the narrow and wide basins. The parameters $A_n$ and $A_w$ denote the areas of the $t^n$ and $t^w$ boxes, respectively.
	
	By volume conservation, changes in the pycnocline volume are mirrored by opposite changes in the deep box volumes, i.e.
	\begin{align*}
		\frac{\partial V_d^i}{\partial t}=-\left(\frac{L_y L_x^i}{2}+A_i\right)\frac{\partial D_i}{\partial t},
	\end{align*}
	where $V_d^i$ represents the deep volume box in basin $i\in\{n,w\}$.
	
	The volume flux from the southern box into $t_s^n$ and $t_s^w$ box is given by $\psi_s^n$ and $\psi_s^w$. This volume flux contains the combined effect of the Eulerian Ekman ($\psi_{ek}^{i}$) and eddy-driven ($\psi_e^{i}$) volume transport, i.e. the residual transport and is computed as:
	\begin{equation}
		\psi_s^i=\psi_{ek}^i-\psi^i_e
		=\frac{\tau L^i_x}{\rho_0|f_s|}-\kappa_{gm}\frac{L_{x}^i}{L_y}D_i.
		\label{eq:psi_s}
	\end{equation}
	Here, $\tau$ denotes the zonal wind-stress strength, $\kappa_{gm}$ the eddy thickness diffusivity \citep{gent1990isopycnal}, $\rho_0$ a reference density, and $f_s$ the Coriolis parameter at $50^\circ$S. We note that $\psi_s^i$ can be positive or negative, depending on whether Eulerian wind-driven upwelling or eddy-driven downwelling dominates, respectively—hence the double arrows in Fig.~\ref{fig:Model_Schematic}a.
	
	The strength of diffusive upwelling into the $t^i$ box follows from the scaled vertically integrated advection–diffusion balance \citep{welander1959advective}, yielding
	\begin{equation}
		\psi_u^i = A_i \frac{\kappa_v}{D_i},
		\label{eq:psi_u}
	\end{equation}
	where $\kappa_v$ denotes the vertical diffusivity.
	
	For strong mixing, basin sinking is no longer confined to a single hemisphere, and the pole-to-pole character of the upper ROC branch disappears \citep{vanderborght2025reduced}. The \cite{gnanadesikan1999simple} model is therefore only an adequate conceptualization of the upper pole-to-pole ROC branch in the quasi-adiabatic limit (i.e., the low-mixing limit; \cite{wolfe2011adiabatic}). In this limit, the pole-to-pole overturning strength scales with the range of overlapping isopycnals between the Southern Ocean and the deep-water formation region in the Northern Hemisphere \citep{nikurashin2012theory}. This range can be quantified as the difference between the densest outcropping water in the Northern Hemisphere and the lightest outcropping water in the Southern Ocean adiabatic upwelling region, roughly at $50^\circ$S (Fig.~\ref{fig:Model_Schematic}b). 
	
	Based on this physical argument, \cite{cimatoribus2014meridional} proposed that the sinking strength in the $n^i$ box follows
	\begin{equation}
		\psi_n^i = \eta \frac{\rho_n^i - \rho_{t_s}^i}{\rho_0} D_i^2 \mathcal{H}(\rho_n^i - \rho_{t_s}^i),
		\label{eq:qni}
	\end{equation}
	where $\eta$ is a hydraulic constant, $\rho_n^i$ and $\rho_{t_s}^i$ denote the densities in the $n^i$ and $t_s^i$ boxes, respectively, and $\mathcal{H}$ is the Heaviside function. The Heaviside function ensures that no pole-to-pole overturning circulation can be sustained in the absence of overlapping isopycnals between the two outcropping regions.
	
	Following \cite{jones2016interbasin} and \cite{ferrari2017model}, we consider an interbasin geostrophic transport, $\psi_g$, arising from the difference in geostrophic zonal flow between the two sectors of the circumpolar domain. When positive, the upper-layer volume transport is directed from the wide into the narrow basin (Fig.~\ref{fig:Model_Schematic}b). Its strength is given by
	\begin{equation}
		\psi_g = \frac{g'}{2 |f_s|} \left(D_w^2 - D_n^2\right),
		\label{eq:psig}
	\end{equation}
	where $g'$ is the reduced gravity in the two-layer system, taken to be constant in this study.
	\subsection{Salinity Equations}\label{BM_Form_Salt}
	The densities $\rho_n^i$ and $\rho_{t_s}^i$ are given by the linear equation of state $\rho = \rho_0 \left(1 - \alpha (T - T_0) + \beta (S - S_0)\right)$, where $\alpha$ and $\beta$ denote the constant thermal expansion and haline contraction coefficients, and $T_0$ and $S_0$ are reference values of temperature and salinity. In \cite{jones2016interbasin}, the density difference $\rho_n^i - \rho_{t_s}^i$ is assumed constant. Here, we adopt a similar assumption for the temperature contrast $T_n - T_{t_s}$, which is taken to be identical in both basins, justified by the relatively fast relaxation timescales of temperature. In contrast, the salinities $S_n^i$ and $S_{t_s}^i$ are treated as dynamic variables. Consequently, prognostic equations for the salt content (i.e., salinity multiplied by box volume) are required for each box.
	
	The salt content of each box may change due to advective overturning fluxes, surface freshwater fluxes, and gyre exchanges, which are physically advective but here represented as a diffusive mixing process (Fig.~\ref{fig:Model_Schematic}a). We consider two types of surface freshwater fluxes: a hemispherically symmetric flux, $E_s^i$, which we here interpret as the hydrological cycle intensity, representing moisture transport from the $t^i$-box into the high-latitude $n^i$ and $s$ boxes; and an interbasin flux, $E_{ib}$, taken as positive when net moisture transport occurs from the northern narrow ($n^n$) to the northern wide box ($n^w$).
	
	The salt-content equations for each box are provided in Appendix~A. The equations governing salt evolution in the $s$-box, the pycnocline boxes, and the $d^i$- and $n^i$-boxes of the two-basin model are largely similar to those of the one-basin model of \cite{cimatoribus2014meridional}. The main differences are the inclusion of an interbasin surface freshwater flux, $E_{ib}$, and an interbasin salt transport associated with the geostrophic volume transport. When $\psi_g > 0$, a thermocline salt flux $\psi_g S_t^w$ and a deep-ocean salt flux $\psi_g S_d^n$ are directed from $t^w$ to $t^n$ and from $d^n$ to $d^w$, respectively. Conversely, when $\psi_g < 0$, the transports reverse direction: the thermocline salt flux $|\psi_g| S_t^n$ and the deep-ocean salt flux $|\psi_g| S_d^w$ are directed from $t^n$ to $t^w$ and from $d^w$ to $d^n$.
	
	Our model can therefore represent two commonly recognized upper-level overturning pathways associated with markedly different salt transports into the overturning basin \citep{speich2001warm}. The first is a warm-route salt transport, in which relatively saline thermocline waters are transferred between basins through interbasin geostrophic flow. The second is a cold-route salt transport, whereby relatively fresh $t_s^i$-box waters enter the $t^i$-box. A similar distinction between overturning pathways was made by \cite{wood2019observable}; however, in their framework the relative contributions ($\gamma$ in their work) were prescribed and fixed, which is unlikely to remain realistic when the GOC transitions to a different sinking configuration. In contrast, our model explicitly resolves both pathways, albeit in a highly simplified manner.
	
	
	The model conserves the total salt content, which can be expressed as
	\begin{equation}
		V_{\rm tot} S_0 = V_s S_s+\sum_{i \in \{n,w\}} \sum_{j} V_j^i S_j^i,
		\label{total_salt}
	\end{equation}
	where $S_0$ is the volume-averaged salinity, $V_s$ is the Southern box volume, and $V_{\rm tot}$ is the total ocean volume, given by
	\begin{equation*}
		V_{\rm tot} = V_s+\sum_{i \in \{n,w\}} \sum_{j} V_j^i,
	\end{equation*}
	with $j \in \{n^i, t^i, d^i, t_s^i\}$. Equation~(\ref{total_salt}) therefore states that the volume-weighted salinity of the system must always equal $S_0$. This constraint is enforced by diagnostically evaluating $S_d^w$ from equation~(\ref{total_salt}). As a result, the model is 10-dimensional, consisting of two prognostic equations for the pycnocline depths and 8 prognostic equations for the salt-content evolution.
	
	\subsubsection{Continuation of Equilibria}\label{BM_Form_Cont}
	The model can be written as an autonomous dynamical system of the form
	\begin{equation}
		\frac{d\boldsymbol{u}(t)}{dt}
		= \boldsymbol{G}\bigl(\boldsymbol{u}(t); \boldsymbol{p}\bigr),
		\label{dyn_system}
	\end{equation}
	where $\boldsymbol{u} = (D_n, D_w, S_n^n, S_n^w, S_t^n, S_t^w, S_{ts}^n, S_{ts}^w, S_d^n, S_s)$ is a 10-dimensional state vector,
	$\boldsymbol{p} \in \mathbb{R}^m$ denotes a set of independent parameters, and $\boldsymbol{G} : \mathbb{R}^{10} \times \mathbb{R}^m \rightarrow \mathbb{R}^{10}$ is a nonlinear vector field. We are interested in steady-state solutions $\boldsymbol{u}_0$ of~(\ref{dyn_system}), satisfying $\boldsymbol{G}(\boldsymbol{u}_0; \boldsymbol{p}) = \boldsymbol{0}$. For a given parameter set $\boldsymbol{p}_0$ admitting such a solution, we trace the dependence of $\boldsymbol{u}_0$ on the value of a selected parameter $g \in \boldsymbol{p}$. This continuation allows us to determine steady states as functions of $g$. Using BifurcationKit.jl \citep{veltz:hal-02902346}, we assess the stability of each equilibrium along the branches.
	
	The model parameters, $\boldsymbol{p}$, are listed in Table~\ref{tab:model_params}. We prescribe the wide basin to be twice as wide as the narrow basin and define one unit of basin width as the width of the narrow basin. The wide basin geometry is therefore characterized by $u^w = 2$ units of basin width. For example, $A_n^w = u^w A_n$, and similarly $A_n^n = u^n A_n$, with $u^n = 1$. 
	
	Assuming a zonally uniform wind stress, a Sverdrup balance implies that the wind-driven upper-layer transport is twice as large in the wide basin as in the narrow basin. Hence, the gyre strength per unit basin width is chosen to be identical in both basins, i.e. $r_n^w=u_n^w r_n$, $r_s^w=u_n^wr_s$, $r_n^n=r_n$ and $r_s^n=r_s$. The symmetric freshwater flux $E_s$ per unit basin width is also prescribed to be identical in both basins, such that $E_{ib}$ fully characterizes the zonally asymmetric surface buoyancy forcing.
	\begin{table*}[t]
		\centering
		\caption{Model parameters for the reference solution. Per unit basin width is abbreviated as p.u.b.w. The model parameter value of $\eta$ was tuned to GCM experiments (Section~\ref{S:CCM}).}
		\label{tab:model_params}
		\footnotesize
		\renewcommand{\arraystretch}{1.1}
		\begin{tabular}{|lll | lll|}
			\hline
			Variable & Value & Explanation & Variable & Value & Explanation \\
			\hline			
			$L_y$ & $1.8\times10^6$\,m & Meridional length of $t_s$ box& $L_x$ & $6.5\times10^6$\,m & Zonal domain width p.u.b.w\\
			
			$A_n$ & $5\times10^{13}$\,m$^2$ & Area of $t$-box p.u.b.w & $V_n$ & $3\times10^{15}$\,m$^3$ & Volume of $n$-box p.u.b.w\\
			
			$V_s$ & $9\times10^{15}$\,m$^3$ & Volume of Southern box &	$V_{\rm tot}$ & $2.5\times10^{17}$\,m$^3$ & Basin ocean volume p.u.b.w\\		
			
			$g'$ & $0.004$\,m\,s$^{-2}$ & Reduced gravity & $\rho_0$ & $1035$\,kg\,m$^{-3}$ & Reference seawater density\\
			
			$S_0$ & $35$\,g\,kg$^{-1}$ & Reference salinity & $f_s$ & $1\times10^{-4}$\,s$^{-1}$ & Coriolis parameter at $50^\circ$S\\
			
			$\kappa_{\rm gm}$ & $5\times10^2$\,m$^2$\,s$^{-1}$ & Eddy thickness diffusivity & $\tau$ & $0.1$\,N\,m$^{-2}$ & Zonal wind stress \\ 
			
			$\kappa_v$ & $2\times10^{-5}$\,m$^2$\,s$^{-1}$ & Vertical diffusivity & $\eta$ & $1.5\times10^4$\,m\,s$^{-1}$ & Hydraulic constant\\
			
			$r_n$ & $5\times10^6$\,m$^3$\,s$^{-1}$ & Northern gyre strength p.u.b.w & $r_s$ & $10\times10^6$\,m$^3$\,s$^{-1}$ & Southern gyre strength p.u.b.w\\
			
			$T_n$ & $5^\circ$C & $n$-box temperature &  $T_{t_s}$ & $9^\circ$C & $t_s$-box temperature \\
			
			$E_s$ & $0.32\times 10^6$\,m$^3$\,s${-1}$ & Symmetric surface freshwater flux p.u.b.w & $u^w$ & 2 & units of basin width for wide basin\\
			\hline
		\end{tabular}
	\end{table*}
	\subsection{Reference Bifurcation Diagram}\label{S:RefSol_GCJM}
\begin{figure}
		\captionsetup{justification=centering}
		\centering
		\includegraphics[width=0.5\linewidth]{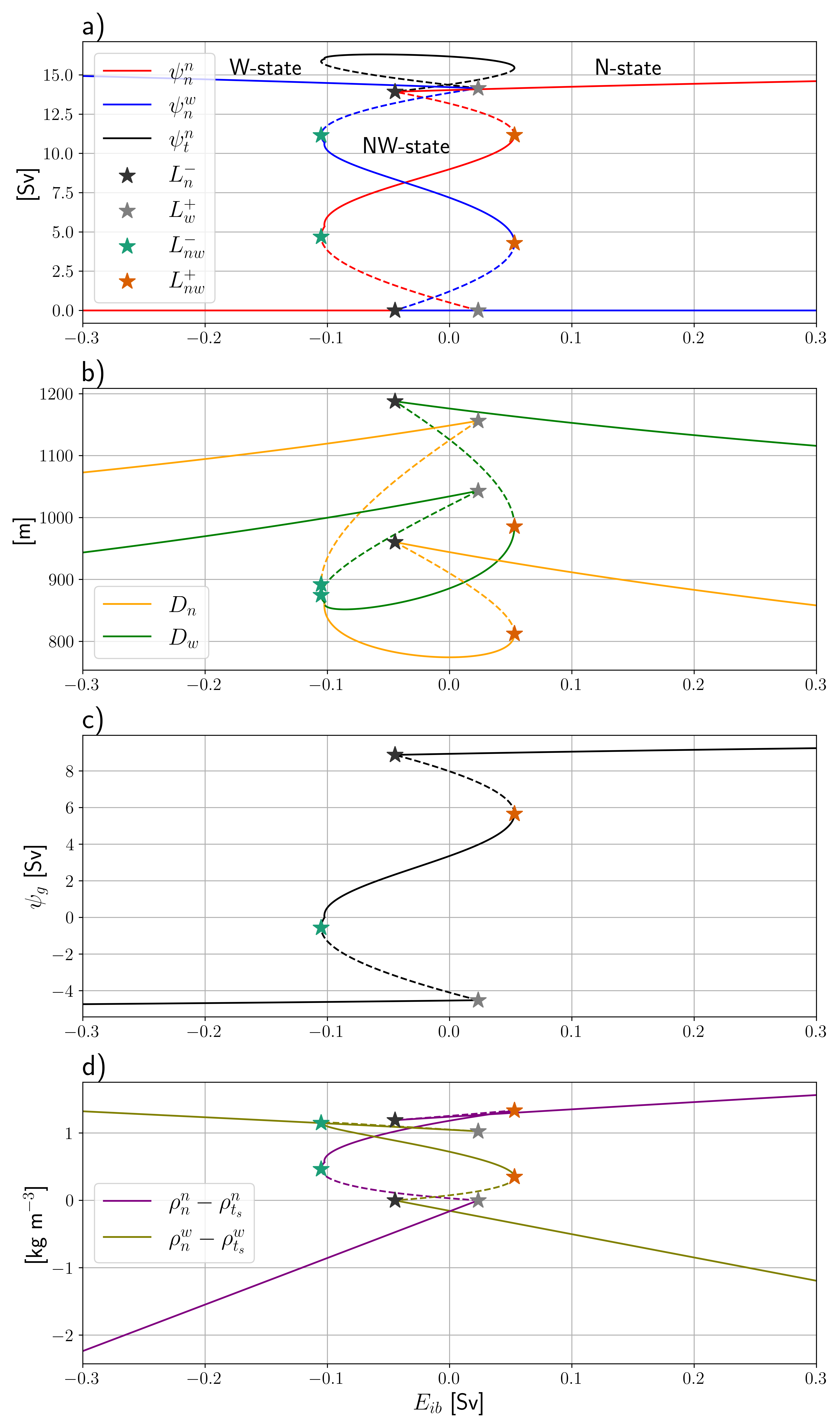}
		\caption{Dependence of (a) $\psi_n^n$ (red), $\psi_n^w$ (blue), and $\psi_n^t$ (black), (b) $D_n$ (green) and $D_w$ (orange), (c) $\psi_g$, and (d) $\rho_n^n - \rho_{t_s}^n$ (purple) and $\rho_n^w - \rho_{t_s}^w$ (olive green) to the interbasin freshwater flux $E_{ib}$, with $E_s^n = 0.32$~Sv. Note that $\psi_n^t$ is shown only for the NW-state. Stable (unstable) equilibria are indicated by thick (dashed) lines. Bifurcation are marked with stars.}
		\label{fig:bif_Eib}
	\end{figure}

	Using the parameter values in Table~\ref{tab:model_params} and continuing all equilibria as a function of the interbasin freshwater flux $E_{ib}$, we obtain the result shown in Fig.~\ref{fig:bif_Eib}. The CTBM admits three possible stable equilibria as a function of $E_{ib}$.
	\begin{enumerate}
		\item An asymmetric sinking state with sinking confined to the wide basin (the W-state), bounded by a single saddle node bifurcation $L_w^+$ and stable for $E_{ib} \in (-\infty, 0.02~\mathrm{Sv}]$.
		\item An asymmetric sinking state with sinking confined to the narrow basin (the N-state), bounded by a single saddle node bifurcation $L_n^-$ and stable for $E_{ib} \in [-0.04~\mathrm{Sv}, +\infty)$.
		\item A symmetric sinking state with active sinking in both basins (the NW-state), bounded by two saddle node bifurcations $L_{nw}^-$ and $L_{nw}^+$ and stable for $E_{ib} \in [-0.1~\mathrm{Sv}, 0.05~\mathrm{Sv}]$.
	\end{enumerate}
	Here, the terms symmetric and asymmetric refer exclusively to the spatial distribution of sinking. The state is termed symmetric when deep-water formation is active in both basins, regardless of differences in its magnitude.
	
	The total sinking strength ($\psi_n^t=\psi_n^w+\psi_n^n$) of the asymmetric overturning states is similar. To first order, it is set by the domain-integrated upwelling, which remains approximately invariant regardless of whether sinking occurs in the narrow or wide basin. The basin with active sinking is characterized by a shallower pycnocline than the passive basin (Fig.~\ref{fig:bif_Eib}b), which from~(\ref{eq:psig}) implies $\psi_g>0$ for the N-state and $\psi_g<0$ for the W-state (Fig.~\ref{fig:bif_Eib}c). In the N-state, this pycnocline depth difference—and hence the magnitude of $\psi_g$—is maximized. This occurs because the volume flux into the wide-basin pycnocline is larger due to its greater area. Consequently, passive-basin upwelling in the N-state exceeds that in the W-state, requiring a larger compensating interbasin transport toward the active basin. These findings are consistent with \cite{jones2016interbasin}.
	
	In the asymmetric overturning states, low salinity in the northern box of the passive basin inhibits sinking, i.e., $\rho_n^i-\rho_{t_s}^i<0$ (Fig.~\ref{fig:bif_Eib}d, equation~(\ref{eq:qni})). For decreasing $E_{ib}$, the density contrast in the passive and active basins of the N-state increases and decreases, respectively. Conversely, for increasing $E_{ib}$, the same behavior occurs in the W-state. This follows from the freshening and salinifying tendencies in the wide- and narrow-basin northern boxes, respectively, induced by increasing $E_{ib}$.
	
	These tendencies in the density contrast explain the weakening of $\psi_n^n$ with increasing $E_{ib}$ in the N-state and of $\psi_n^w$ with decreasing $E_{ib}$ in the W-state (Fig.~\ref{fig:bif_Eib}a). In response, the active-basin pycnocline deepens, reducing the magnitude of $\psi_g$ and causing the passive-basin pycnocline to deepen as well (Fig.~\ref{fig:bif_Eib}b,c).  The pycnocline adjustments therefore damp changes in sinking strength (equation~(\ref{eq:qni})). These dependencies persists until the asymmetric sinking state terminates at the $E_{ib}$ value for which the passive-basin density contrast vanishes (i.e. $\rho_n^i=\rho_{t_s}^i$)—at the $L_n^-$ saddle node bifurcation for the N-state and at $L_w^+$ for the W-state (Fig.~\ref{fig:bif_Eib}d).
	
	The total sinking strength of the NW-state is approximately 2~Sv larger than in the asymmetric states (Fig.~\ref{fig:bif_Eib}a). This arises because the NW-state is characterized by reduced $D_n$ and $D_w$ relative to the asymmetric states (Fig.~\ref{fig:bif_Eib}b). From the steady-state forms of~(\ref{Dn}) and~(\ref{Dw}), the domain-integrated upwelling is consequently larger. The overall shallower pycnocline in the NW-state results from active sinking in both basins, which more efficiently drains the pycnocline boxes.
	
	The sensitivity of $\psi_n^w$ and $\psi_n^n$ to $E_{ib}$ is much stronger in the symmetric overturning than in the asymmetric case. As $E_{ib}$ increases, $\psi_n^w$ decreases while $\psi_n^n$ increases. This seesaw in sinking strengths produces a total overturning that depends only weakly on $E_{ib}$ and reflects a similar pattern in the density contrasts (Fig.~\ref{fig:bif_Eib}d): the density contrast in the narrow basin rises with $E_{ib}$ due to the salinification of the $n^n$ box, whereas it decreases in the wide basin as the $n^w$ box undergoes compensating freshening. 

	In the NW-state, $\psi_g > 0$ for most $E_{ib}$ values, except near $L_{nw}^-$ (Fig.~\ref{fig:bif_Eib}c,d). This behavior results from the basin-size asymmetry: the larger area of the wide-basin pycnocline maintains $D_n < D_w$, although this difference diminishes when $\psi_n^w$ becomes sufficiently larger than $\psi_n^n$ at low $E_{ib}$. As $E_{ib}$ increases, $\psi_g$ grows, with a sensitivity that is markedly stronger than in the asymmetric states.
	
	The enhanced sensitivity of the NW-state arises from two complementary mechanisms. First, because $E_{ib}$ forces changes in the sinking strength of both basins, $\psi_g$ responds more strongly to $E_{ib}$. This enhanced response weakens, and may even reverse, the stabilizing pycnocline feedback that normally counteracts variations in sinking strength. Indeed, Fig.~\ref{fig:bif_Eib}b shows that, in the NW-state, $D_n$ depends only weakly on $\psi_n^n$ and, for some values, even increases as $\psi_n^n$ grows. Second, as $E_{ib}$ increases, the amplified response of $\psi_g$ enhances interbasin salt transport from the wide to the narrow basin. Because $S_n^n < S_t^w$ for all $E_{ib}$ values (not shown), this transport salinifies the narrow basin. This, in turn, reinforces the initial salinity anomalies, the associated sinking responses in both basins, and $\psi_g$. Together, these mechanisms explain the elevated sensitivity of the NW-state to increases in $E_{ib}$.
	
	The N-state is the only stable equilibrium for $E_{ib} > 0.05$~Sv, while the W-state is the only stable equilibrium for $E_{ib} < -0.1$~Sv. In this large-magnitude forcing regime, the asymmetric sinking state is uniquely determined by the imposed interbasin freshwater flux. However, for $E_{ib} \in [-0.1~\mathrm{Sv}, 0.05~\mathrm{Sv}]$, multiple equilibria coexist under identical forcing. This sensitivity to initial conditions arises from the salt-advection feedback, which can destabilize the symmetric overturning state: the circulation redistributes salinity by converging it in one basin and extracting it from the other via interbasin transport, thereby localizing sinking in a single basin \citep{nilsson2013ocean}. Whether and in what basin such localization occurs, depends on the initial state of the system.
	
	The bifurcation diagram in Fig.~\ref{fig:bif_Eib} is asymmetric about $E_{ib}=0$, with the symmetry axis shifted toward negative $E_{ib}$. As a result, asymmetric sinking in the wide basin is stable for a smaller range of $E_{ib}$ than in the narrow basin. Moreover, symmetric overturning remains stable over a narrower interval of positive than negative $E_{ib}$, implying that a smaller interbasin freshwater-flux magnitude is required to uniquely localize sinking in the narrow basin. These features of the bifurcation structure stem from the imposed basin-size asymmetry and indicate that the salt-advection feedback more readily dominates to localize asymmetric sinking in the narrow basin than in the wide basin. The dynamical origin of this preference is examined in the following section.
	
	\subsection{Sensitivity on the Hydrological Cycle}\label{S:Es_Sens}
	Results of the previous section raise questions about how the ability of interbasin freshwater fluxes to uniquely localize sinking depends on key model parameters. The uniqueness of sinking states is controlled by the locations of the bifurcation points $L_n^-$, $L_w^+$, $L_{nw}^-$, and $L_{nw}^+$. Analytical expressions for the $E_{ib}$ value at the bifurcation points $L_n^-$ and $L_w^+$ can be derived by considering the circulation in the adiabatic limit and noting that, at these bifurcation points, the density difference in the passive basin vanishes; that is, $\rho_n^w - \rho_{t_s}^w = 0$ and $\rho_n^n - \rho_{t_s}^n = 0$, respectively (Fig.~\ref{fig:bif_Eib}d). The full derivation is outlined in Appendix~B, and yields:
	\begin{subequations}
		\begin{align}
			E_{ib}(L_n^-)&=u^w E_s \left(\frac{r_n^w-r_s^w-\psi_g}{\psi_g+r_s^w+r_n^w}\right)
			+\frac{\alpha \Delta T r_n^w}{\beta S_0}
			\left(\frac{r_s^w+\psi_g}{\psi_g+r_s^w+r_n^w}\right)
			\label{Lnmin_adiabatic},\\
			E_{ib}(L_w^+)&=E_s\left(\frac{r_s^n-\psi_g-r_n^n}{r_s^n+r_n^n-\psi_g}\right)-\frac{\alpha \Delta T r_n^n}{\beta S_0}\left(\frac{r_s^n-\psi_g}{r_s^n+r_n^n-\psi_g}\right)
			\label{Lwmax_adiabatic}.
		\end{align}
		\label{eq:saddle_dependence}
	\end{subequations}
	Here $E_{ib}(L_n^-)$ and $E_{ib}(L_w^+)$ are the $E_{ib}$ values at the bifurcation points $L_n^-$ and $L_w^+$, respectively. We defined $\Delta T \equiv T_{t_s} - T_n$. Note that, apart from $\psi_g$, all quantities appearing in equation~(\ref{Lnmin_adiabatic}) are prescribed parameters of the autonomous dynamical system. Moreover, because $\psi_g$ at $L_n^-$ and $L_w^-$ weakly depends on parameters that do not modulate the domain integrated upwelling, we expect it to remain approximately constant when changing the parameters in~(\ref{eq:saddle_dependence}). 
	
	Similar expressions for $E_{ib}(L_{nw}^-)$ and $E_{ib}(L_{nw}^+)$ can, in principle, be derived, but they are algebraically cumbersome and provide limited physical insight. We therefore describe their dependence on the key model parameters qualitatively.
	%
	%
	%
	
\begin{figure}
		\captionsetup{justification=centering}
		\centering
		\includegraphics[width=0.5\linewidth]{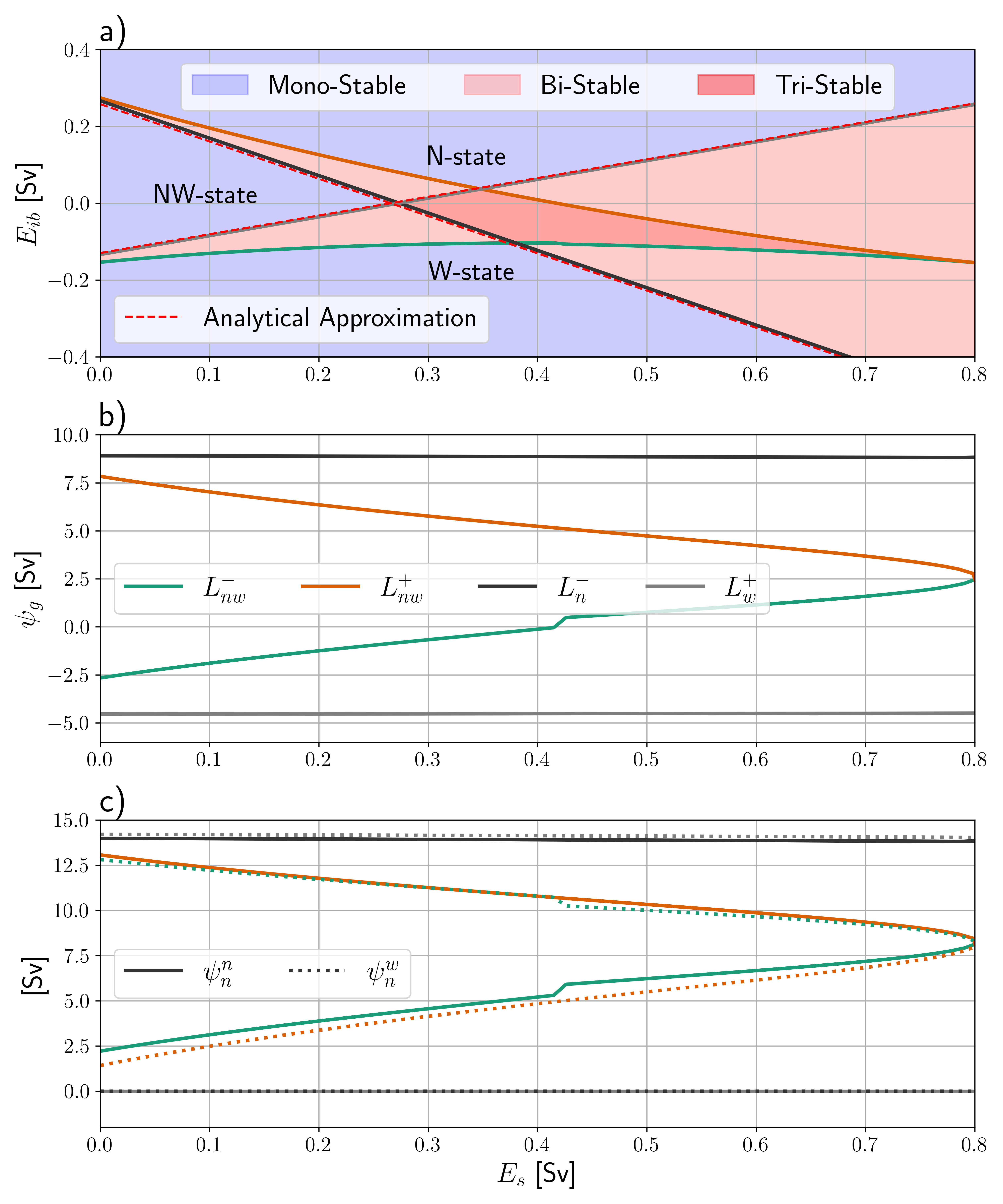}
		\caption{(a) Interbasin surface freshwater flux, (b) interbasin geostrophic transport and (c) sinking strength ($\psi_n^n$: thick and $\psi_n^w$: dotted), on saddle node bifurcations $L_n^-$, $L_w^+$, $L_{nw}^-$ and $L_{nw}^+$ versus $E_s$. Dashed red lines in (a) correspond to analytical expressions~(\ref{Lnmin_adiabatic}) and~(\ref{Lwmax_adiabatic}). In (a), the parameter space is separated into a mono-, bi- and tri-stable regime.}
		\label{fig:cont_saddle_Esa}
	\end{figure}

	%
	%
	Fig.~\ref{fig:cont_saddle_Esa}a shows the dependence of the $E_{ib}$ value at the saddle points on $E_s$. The approximations~(\ref{Lnmin_adiabatic}) and~(\ref{Lwmax_adiabatic}) closely reproduce the behavior of $E_{ib}(L_n^-)$ and $E_{ib}(L_w^+)$, indicating that the diffusive component of the circulation plays little role in determining the saddle-point locations. The saddle points shift linearly toward lower and higher $E_{ib}$, respectively. The linear dependence of the saddle-node $E_{ib}$ value, on $E_s$, indicates that $\psi_g(L_n^-)$ and $\psi_g(L_w^+)$ is approximately invariant with respect to $E_s$, as verified in Fig.~\ref{fig:cont_saddle_Esa}b. This weak sensitivity follows from the relatively weak dependence of the asymmetric sinking strength on $E_s$ and $E_{ib}$ (Fig.~\ref{fig:cont_saddle_Esa}c).
	
	The shift of $L_n^-$ and $L_w^+$ toward lower and higher $E_{ib}$, respectively, follows from~(\ref{Lnmin_adiabatic}) and~(\ref{Lwmax_adiabatic}), where the coefficients satisfy $r_n^w - r_s^w - \psi_g < 0$ and $r_s^n - \psi_g - r_n^n > 0$. Consequently, as $E_s$ increases, the range of $E_{ib}$ over which the N- and W-states are stable widens.
	Physically, this follows from the fact that a larger $E_s$ increases the surface freshwater flux toward northern box, maintaining a negative density contrast, in the passive basin, over a broader range of $E_{ib}$-values. 
	
	However, as indicated by equations~(\ref{Lnmin_adiabatic})-(\ref{Lwmax_adiabatic}), the effectiveness of the $E_s$ induced freshening of the passive basin northern box, i.e. the slope in Fig.~\ref{fig:cont_saddle_Esa}a, is modulated by salt transport processes within the thermocline of the passive basin. In particular, these include salt transport by the northern and southern gyre, and salt transport associated with the geostrophic exchange flow.
	
	Interestingly, $\psi_g$ in the N-state approximately equals $-u^w\psi_g$ in the W-state (Fig.~\ref{fig:cont_saddle_Esa}b). Using this relation in~(\ref{eq:saddle_dependence}) yields $E_{ib}(L_n^-) = -u^w E_{ib}(L_w^+)$, implying that the N-state loses stability for values of $E_{ib}$ that are twice as large and of opposite sign compared to the W-state. Consequently, when $E_{ib}(L_n^-) < 0$, the range of $E_{ib}$ values for which the N-state is stable is larger than that of the W-state and vice-versa for $E_{ib}(L_n^-) > 0$. The explanation is straightforward: because the northern box of the narrow basin is half as wide as the corresponding box in the wide basin, while both basins have the same gyre strength per unit basin width. As a result, a given interbasin freshwater flux $E_{ib}$ induces a freshening or salinification tendency in the passive basin that is twice as strong in the narrow basin. Consequently, the critical condition $\rho_n^i - \rho_{t_s}^i = 0$ is reached for smaller magnitudes of $E_{ib}$ in the W-state than in the N-state. 
	
	Changes in the NW-state with increasing $E_s$ exhibit two features: (1) the stability range in $E_{ib}$ shrinks, and (2) the state shifts toward lower $E_{ib}$. These effects follow from $L_{nw}^+$ shifting toward lower $E_{ib}$ and $L_{nw}^-$ initially shifting toward higher $E_{ib}$ for $\psi_g>0$, but toward lower $E_{ib}$ when $\psi_g<0$ (Fig.~\ref{fig:cont_saddle_Esa}b). For $E_s \approx 0.8$~Sv, the NW-state disappears in a cusp bifurcation. As the state shifts toward lower $E_{ib}$, the entire bifurcation structure lies in the $E_{ib}<0$ regime for $E_s \gtrsim 0.42$~Sv. Thus, beyond this value, symmetric overturning exists only when the wide basin is more evaporative compared to the narrow basin.
	
	To explain these features, we reiterate that the symmetric overturning state loses stability at the saddle points due to the salt-advection feedback. For example, a positive perturbation in $E_{ib}$ weakens $\psi_n^w$ and strengthens $\psi_n^n$, thereby reducing and enhancing the salt influx into the respective northern boxes. Enhanced interbasin salt transport from the wide to the narrow basin amplifies this response. At $L_{nw}^+$, wide-basin overturning reaches a minimum strength. Beyond this point, the temperature contrast can no longer sustain $\rho_n^w-\rho_{t_s}^w>0$, and salt-advection localizes sinking in a single basin. 
	
	With increasing $E_s$, the salinity transport into the northern boxes—and hence the salt-advection feedback—strengthens. This raises the minimum overturning strength (i.e., $\psi_n^n$ at $L_{nw}^-$ and $\psi_n^w$ at $L_{nw}^+$; Fig.~\ref{fig:cont_saddle_Esa}c) above which this feedback dominates. For $E_s \gtrsim 0.8$~Sv, the salt-advection feedback dominates for all $E_{ib}$, and the NW-state disappears. Based on this increase in minimum overturning strength, one might expect $E_{ib}(L_{nw}^-)$ to increase and $E_{ib}(L_{nw}^+)$ to decrease. However, Fig.~\ref{fig:Eib_three_Esas} shows that these shifts contribute only weakly to the overall dependence of $E_{ib}(L_{nw}^-)$ and, in particular, $E_{ib}(L_{nw}^+)$ on $E_s$. Instead, the dominant effect arises from the dependence of the sinking strength on $E_s$ at fixed $E_{ib}$.
	\begin{figure}
		\captionsetup{justification=centering}
		\centering
		\includegraphics[width=0.5\linewidth]{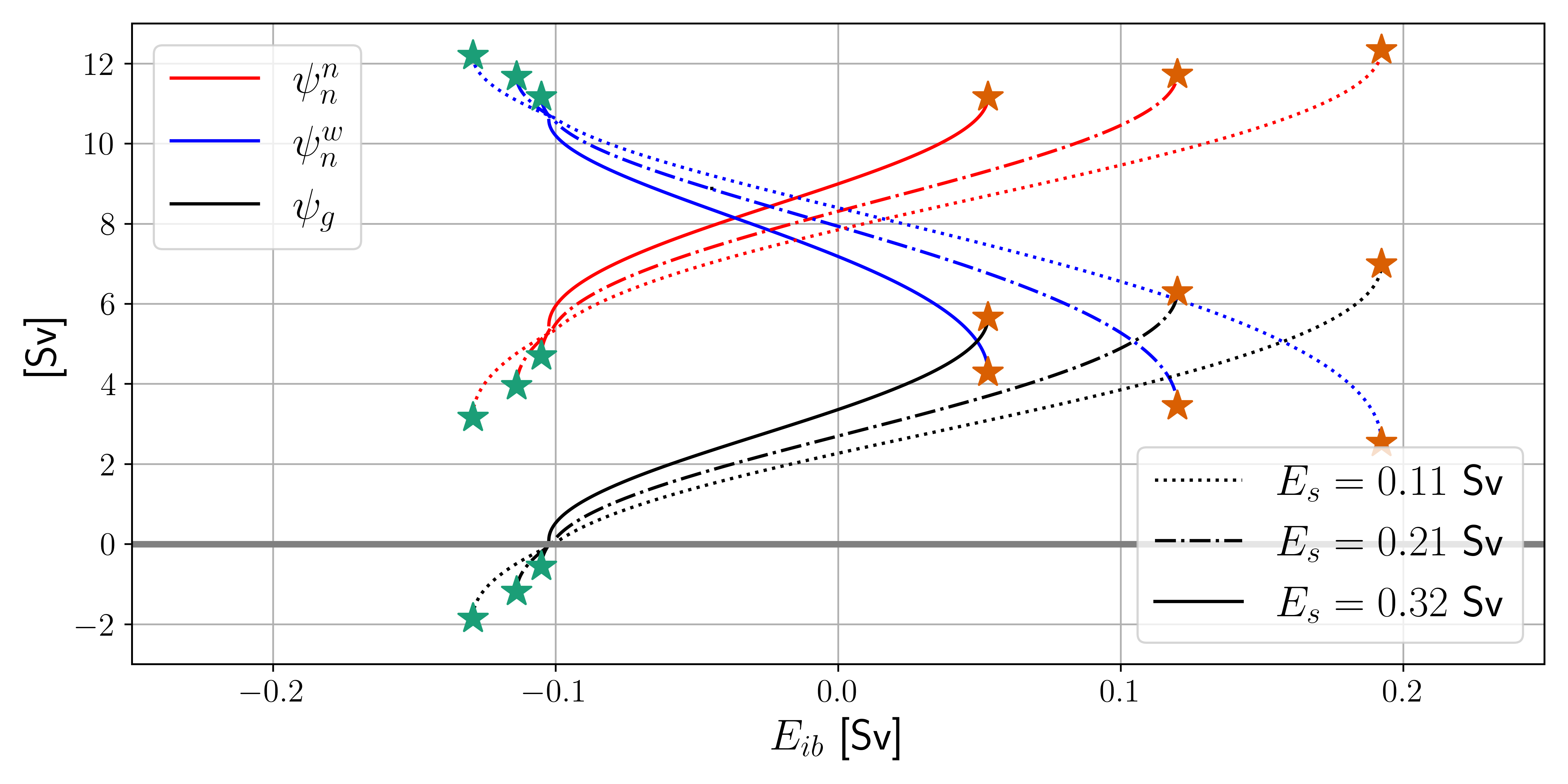}
		\caption{NW-state bifurcation diagram for three different $E_s$. The stars again represent the positions of the saddle node bifurcations.}
		\label{fig:Eib_three_Esas}
	\end{figure}
	
	This is illustrated by continuing the NW-state in $E_s$ at $E_{ib}=0$, noting that it exists only up to $E_s \lesssim 0.42$~Sv (Fig.~\ref{fig:cont_saddle_Esa}a). Fig~\ref{fig:cont_Esa_Eib_00}a shows that, at $E_{ib}=0$, $\psi_n^n$ increases while $\psi_n^w$ decreases with increasing $E_s$, indicating a shift of the NW-state toward negative $E_{ib}$, consistent with Fig.~\ref{fig:Eib_three_Esas}. These changes are driven by corresponding variations in the density contrasts of the two basins (Fig.~\ref{fig:cont_Esa_Eib_00}b). Although increasing $E_s$ enhances the freshwater flux over the northern boxes of both basins, the salinity decrease in the narrow northern box is partly offset by enhanced interbasin salt transport from the wide-basin thermocline—where evaporation increases with $E_s$—to the narrow-basin thermocline ($\psi_g>0$, Fig.~\ref{fig:cont_Esa_Eib_00}c). Consequently, for $\psi_g>0$, increasing $E_s$ strengthens (weakens) sinking in the narrow (wide) basin, whereas the opposite holds when $\psi_g<0$ (Fig.~\ref{fig:Eib_three_Esas}). Moreover, Fig.~\ref{fig:Eib_three_Esas} shows that the magnitude of the NW-state shift with increasing $E_s$ scales with $|\psi_g|$, and vanishes when $\psi_g=0$.
	
	Because $\psi_g$ increases with $E_{ib}$, combined with the increasing minimum overturning strength, this effect leads to a contraction of the $E_{ib}$ range over which the NW-state remains stable as $E_s$ grows (Fig.~\ref{fig:Eib_three_Esas}). Moreover, due to basin-size asymmetry, $\psi_g>0$ for most $E_{ib}$, producing an additional net shift of the NW-state toward negative $E_{ib}$. Thus, interbasin salt transport, and its dependence on $E_{ib}$, explain feature 1 and 2 in Fig.~\ref{fig:cont_saddle_Esa}a. Moreover, the dominance of this mechanism over shifts induced by changes in the minimum overturning strength explains why $L_{nw}^-$ shifts to higher (lower) $E_{ib}$ when $\psi_g<0$ ($\psi_g>0$; Fig.~\ref{fig:cont_saddle_Esa}b).
	\begin{figure}
		\captionsetup{justification=centering}
		\centering
		\includegraphics[width=0.5\linewidth]{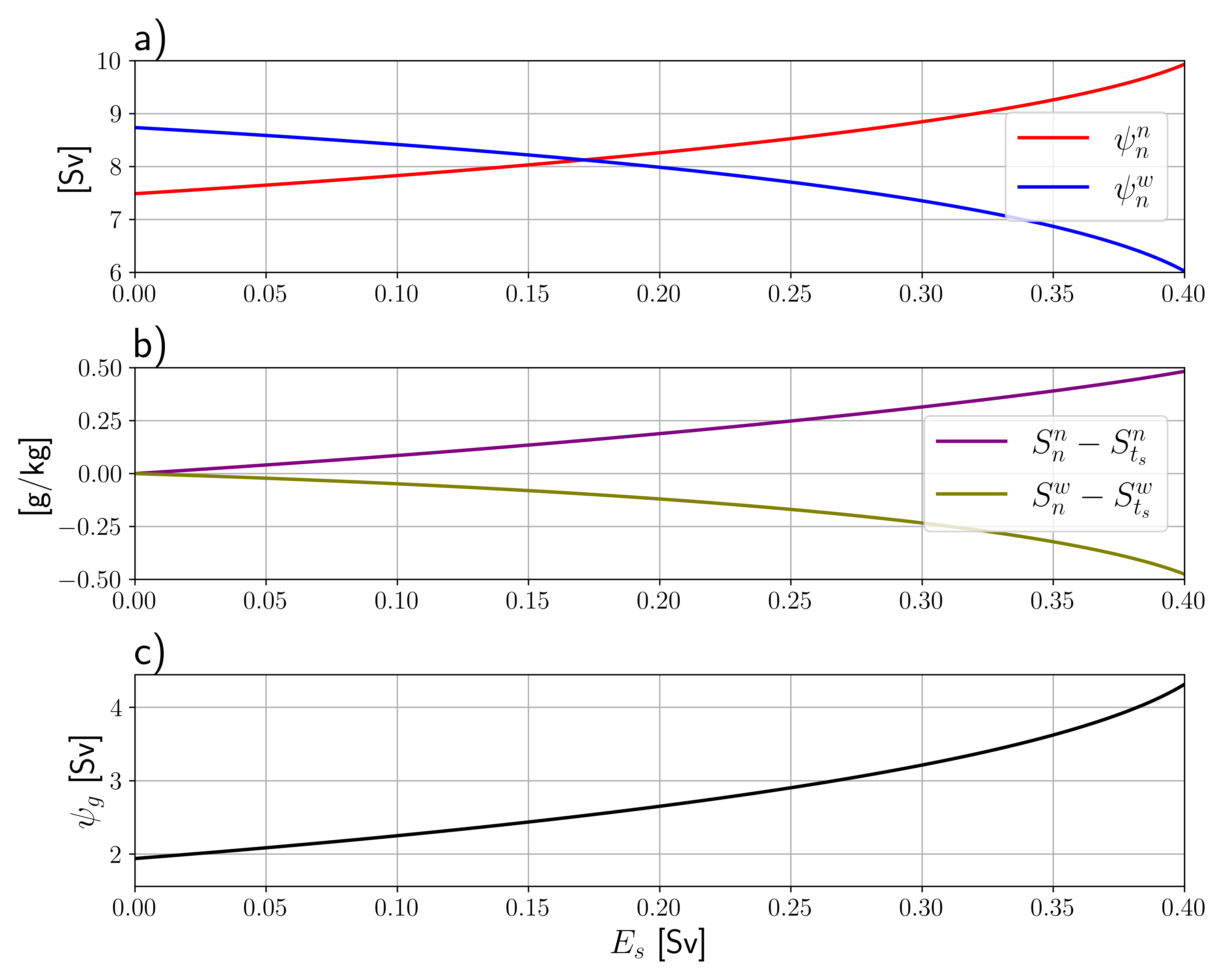}
		\caption{Continuation of (a) sinking strength, (b) salinity contrasts and (c) pycnocline depths, at $E_{ib}=0$~Sv for $E_s$.}
		\label{fig:cont_Esa_Eib_00}
	\end{figure}
	
	In the context of the present-day climate, which is best described by an N-state (i.e., AMOC-like circulation), the CTBM suggests that achieving this state independently of initial conditions requires a large interbasin freshwater flux at low $E_s$ ($\lesssim0.2$~Sv), because the NW-state is monostable over a broad range of $E_{ib}$. As $E_s$ increases, the minimum $E_{ib}$ required to obtain a monostable N-state decreases, as the NW-state both shrinks and shifts toward lower $E_{ib}$. However, around $E_s \approx 0.35$~Sv, $E_{ib}(L_{nw}^+)$ and $E_{ib}(L_w^+)$ coincide. Beyond this point, the minimum $E_{ib}$ required for a monostable N-state increases again, as the stability range of the W-state expands into the positive $E_{ib}$ regime. Thus, under a very strong hydrological cycle, obtaining an N-state independently of initial conditions once more requires large-magnitude $E_{ib}$.
	
	A similar but opposite dependence occurs for the W-state. However, the dynamical landscape is clearly asymmetric about $E_{ib}=0$~Sv (Fig.~\ref{fig:cont_saddle_Esa}a). At low $E_s$, localizing sinking in the wide basin requires smaller $E_{ib}$ magnitudes than in the N-state ($|E_{ib}(L_{nw}^+)| > |E_{ib}(L_{nw}^-)|$). This is because, in the NW-state at low $E_s$, salinity contrasts are weak and the density difference between the narrow and wide basins is primarily set by temperature. Therefore, at $E_{ib}=0$~Sv, sinking is stronger in the wide basin due to its larger upwelling, i.e. $D_w>D_n$ (Fig.~\ref{fig:Eib_three_Esas}; $E_s=0.11$~Sv). 
	
	At intermediate $E_s$ ($0.2 \lesssim E_s \lesssim 0.4$~Sv), a monostable N-state occurs at smaller $E_{ib}$ magnitudes than the W-state ($|E_{ib}(L_{nw}^+)| < |E_{ib}(L_{nw}^-)|$), resulting from the faster decrease of $E_{ib}(L_{nw}^+)$ compared to the increase of $E_{ib}(L_{nw}^-)$, which reflects the larger $\psi_g$ magnitude at $L_{nw}^+$ relative to $L_{nw}^-$ (Fig.~\ref{fig:cont_saddle_Esa}b). At high $E_s$ ($\gtrsim 0.4$~Sv), the $E_{ib}$ magnitude required for a monostable W-state increases more rapidly with $E_s$ than that for the N-state, which follows from $E_{ib}(L_n^-) = -u^w E_{ib}(L_w^+)$.
	
	Thus, in short; the CTBM suggests that for low $E_s$, a PMOC is uniquely attained at smaller $E_{ib}$ magnitudes, whereas for intermediate and high $E_s$, an AMOC is more readily established by interbasin freshwater flux asymmetries.
	\section{Comparison with the MITgcm}\label{S:CCM}
	Results from the CTBM suggest that the magnitude of the interbasin freshwater flux required to break overturning symmetry and localize sinking in a single basin depends on the intensity of the hydrological cycle. Here we test these findings using experiments performed with the Massachusetts Institute of Technology general circulation model (MITgcm; \cite{marshall1997finite, marshall1997hydrostatic}).
	\subsection{Circulation Model and Diagnostics}\label{S:CM}
	The MITgcm is configured following \cite{jones2016interbasin}, on a domain extending from $0$°E to $200$°E in longitude at $2^\circ$ resolution, and from $70$°S to $70$°N in latitude at a resolution of $2^\circ\cos(\theta)$, where $\theta$ denotes latitude. The ocean has a flat bottom at $3800$~m, discretized with 30 unevenly spaced vertical levels, varying from $20$~m at the surface to $200$~m at depth. The domain is divided into two basins, one twice as wide as the other, seperated by vertical walls extending the full depth, located at $67$°E and $220$°E from $70$°N to $50$°S. The basins are connected through a zonally periodic re-entrant channel spanning $70$°S to $50$°S. This channel includes a sill at $67^\circ$E, extending from the ocean bottom to 1300 m above it. The exact longitude of the sill has no impact on the results.
	
	Vertical and horizontal Laplacian viscosities are set to $1\times10^{-4}$~m$^2$~s$^{-1}$ and $5\times10^{4}$~m$^2$~s$^{-1}$, respectively. Along-isopycnal mixing by mesoscale eddies is represented using the Redi scheme \citep{redi1982oceanic}, while their advective contribution is parameterized using the Gent–McWilliams (GM) scheme \citep{gent1990isopycnal}. Both parameterizations use the same mixing coefficient, $\kappa_{gm}=500$~m$^2$~s$^{-1}$. The Redi tensor is exponentially tapered toward horizontal diffusion in regions of weak stratification \citep{danabasoglu1995sensitivity}. The GM scheme is implemented using the boundary-value-problem formulation of \citet{ferrari2010boundary}, with vertical mode number $m=2$ and a minimum wave speed of $c_{\mathrm{min}}=0.1$~m~s$^{-1}$. 
	
	Temperature and salinity are related to buoyancy ($b$) through a linear equation of state,
	\begin{equation*}
		b=g\left[\alpha T-\beta\left(S-S_0\right)\right].
	\end{equation*}
	Temperature and salinity are advected using the flux-limited version of the second-order moment scheme of \citet{prather1986numerical}.  Vertical tracer mixing is represented by vertical diffusion, with enhanced diffusivity near the surface to represent an idealized mixed layer of depth $d=20$~m. The vertical structure of the diffusivity is prescribed as
	\begin{equation*}
		\kappa_v(z)=\kappa_v^b+\kappa_v^m\left[1+\tanh\left(\frac{z+d}{d}\right)\right],
	\end{equation*}
	where $\kappa_v^b=2\times10^{-5}$~m$^2$~s$^{-1}$ is the background vertical diffusivity and $\kappa_v^m=10^{-2}$~m$^2$~s$^{-1}$ characterizes enhanced mixing within the mixed layer. Convective mixing is included by increasing the vertical diffusivity to $100$~m$^2$ s$^{-1}$ in regions with an unstable stratification.
	
	In the surface grid cell, temperature is relaxed toward a prescribed profile $T^*$ with a restoring timescale of $10$~days. The zonally independent target temperature profile reads
	\begin{equation}
		T^*(\theta)=\frac{\Delta T}{2}\left[\cos\left(\frac{\theta}{\theta_n}\right)+1\right]+T_n\exp\left(\frac{-(\theta-\theta_n)^2}{\delta_T^2}\right)
		\label{eq:Tstar}
	\end{equation}
	Here, $\Delta T=25^\circ$C is the equator-to-pole temperature difference, $\theta_n=70^\circ$N, $T_n=1^\circ$C is the minimum temperature in the Northern Hemisphere, and $\delta_T=18^\circ$.
	
	At the surface grid cell, salinity is forced through a flux boundary condition specified by a prescribed freshwater flux,
	\begin{equation}
		F(\theta)=(P-E)_{\mathrm{sym}}(\theta)-\langle(P-E)_{\mathrm{sym}}\rangle
		\label{eq:fw_flux_MIT}
	\end{equation}
	where $(P-E)_{\mathrm{sym}}$ is the longitudinally symmetric component of the surface freshwater flux, and $\langle (P-E)_{\mathrm{sym}} \rangle$ is its domain mean, subtracted to ensure zero net freshwater input. 
	
	The longitudinally symmetric component of the freshwater flux reads \citep{wolfe2014salt}
	\begin{equation*}
		(P-E)_{\mathrm{sym}}(\theta)=-F_0\left[\cos\left(\pi\frac{\theta}{\theta_n}\right)-a_s\exp\left(\frac{\theta^2}{\delta_S^2}\right)\right],
		\label{eq:P_E}
	\end{equation*}
	where $F_0$ is the amplitude of the symmetric surface freshwater flux, $\delta_s=8^\circ$ and $a_s=2$ representing the equatorial enhancement of precipitation.
	
	For comparison with the CTBM, we diagnose the value of $E_s$ from equation~(\ref{eq:fw_flux_MIT}) by integrating
	\begin{equation}
		E_s = a^2 \int_{\lambda_1^n}^{\lambda_2^n} \int_{-\theta_f}^{\theta_f} F(\theta)\cos(\theta) d\theta d\lambda,
		\label{eq:Es_converter}
	\end{equation}
	where $a$ is the Earth's radius, $\lambda$ denotes longitude, and $\lambda_1^n$ and $\lambda_2^n$ define the western and eastern boundary longitudes of the narrow basin, respectively. The latitude $\theta_f$ marks the northernmost latitude at which $F(\theta) = 0$.
	
	The zonal momentum equation is forced at the surface grid cell by a zonal surface wind stress, prescribed as a longitudinally symmetric profile:
	\begin{equation}
		\tau_x(\theta)=\tau_0\left[-\cos\left(\frac{3\pi}{2}\frac{\theta}{\theta_n}\right)+\exp\left(-\frac{\theta^2}{\delta_\tau^2}\right)\right],
		\label{eq:tau_wind}
	\end{equation}
	where $\tau_0 = 0.1$ N m$^{-2}$ and $\delta_\tau = 10^\circ$ \citep{jones2016interbasin}.

\begin{figure}
		\centering
		\includegraphics[width=0.5\linewidth]{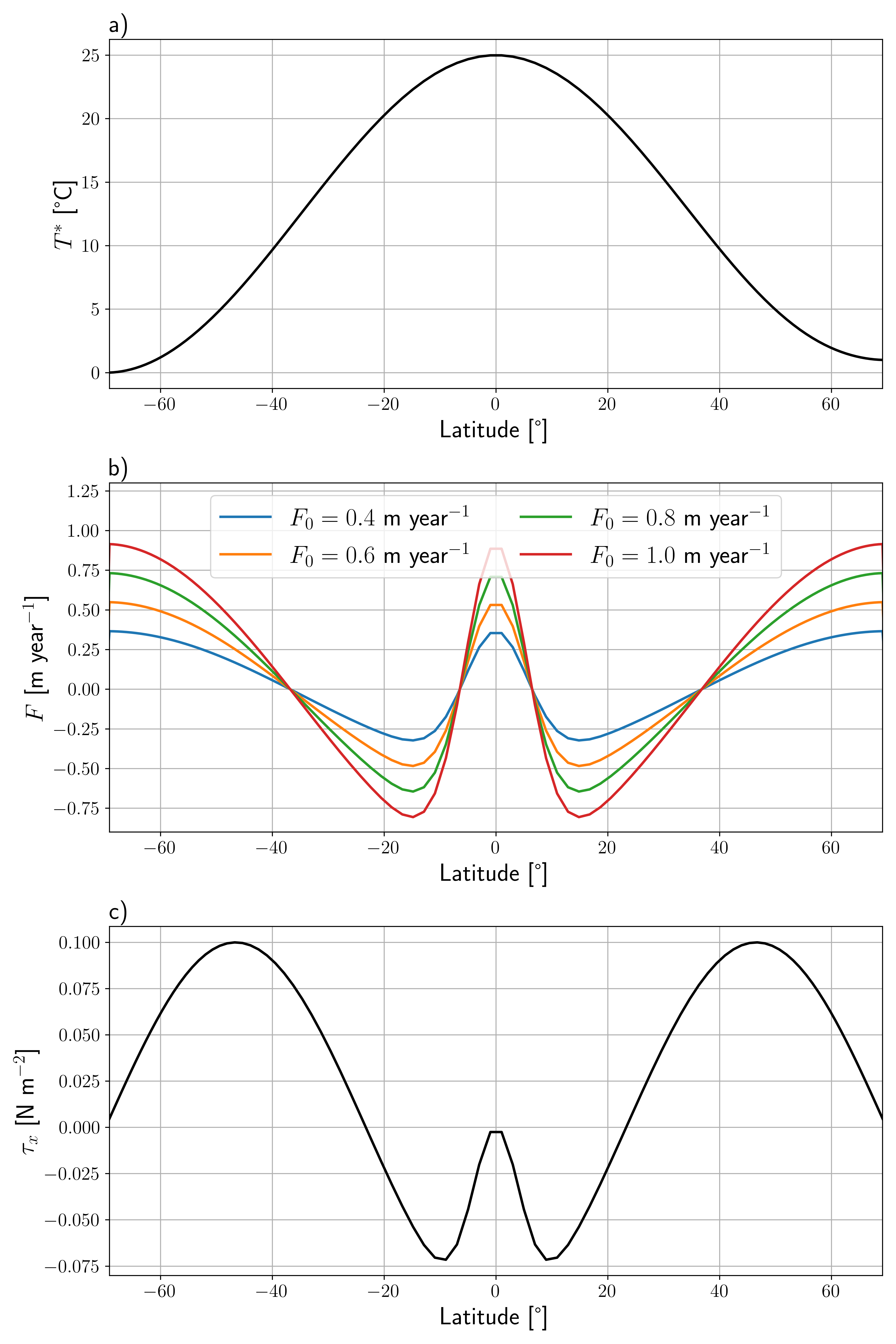}
		\caption{(a) Surface temperature relaxation profile~(\ref{eq:Tstar}). (b) Surface freshwater flux for four different values of $F_0$~(\ref{eq:fw_flux_MIT}). (c) Surface zonal wind-stress profile~(\ref{eq:tau_wind})}
		\label{fig:MIT_surf_forcing}
	\end{figure}

	All surface forcing profiles are shown in Fig.~\ref{fig:MIT_surf_forcing}. Unless stated otherwise, all experiments are initialized from rest with a uniform salinity $S_0=35$ g kg$^{-1}$ and temperature of $0^\circ$C. Each simulation is run for 5000 years, which was verified to be sufficient for the model to reach an approximate statistical steady state in all experiments.
	
	The overturning circulation is diagnosed by calculating the zonally integrated residual overturning streamfunction $\psi^i(\theta,\tilde{b})$ \citep{wolfe2011adiabatic}: 
	\begin{equation}
		\psi^i(\theta, \tilde{b})=a\cos(\theta)\int_{\lambda^i_1}^{\lambda^i_2}\int_{-H}^0 v^\dagger\mathcal{H}[b(\lambda, \theta, z, t)-\tilde{b}]dz d\lambda. 
		\label{eq:psi_res}
	\end{equation}
	With $\lambda_1^i$ and $\lambda_2^i$ represent the western and eastern boundary coordinate of the narrow $i=n$ or wide basin $i=w$.  Within the latitude range of the re-entrant channel, the integral is taken over all longitudes, so that $\psi^n=\psi^w$ within this range. The residual meridional velocity $v^\dagger=v+v_{gm}$ represents the combined effect of the Eulerian and GM bolus velocity field. Hence, equation~(\ref{eq:psi_res}) quantifies the residual transport above a given buoyancy level $\tilde{b}$. For comparison with the CTBM, we define the reference buoyancy level $\hat{b}$ as the buoyancy level at which the overturning streamfunction attains its maximum at the equator over both basins. The northern sinking strengths are then evaluated as $\psi_n^w = \psi^w( 0^\circ,\hat{b})$ and $\psi_n^n = \psi^n(0^\circ,\hat{b})$.
	
	The overturning streamfunction $\psi^i$ can be mapped back onto height coordinates by calculating the zonal mean depth $\mathcal{Z}^i$ of a given buoyancy level $\tilde{b}$:
	\begin{equation}
		\mathcal{Z}^i(\theta,\tilde{b})=\frac{1}{\lambda_2^i-\lambda_1^i}\int_{\lambda_1^i}^{\lambda_2^i}\int_{-H}^0 \mathcal{H}[b(\lambda,\theta,z,t)-\tilde{b}] dz d\lambda.
	\end{equation}
	As in~(\ref{eq:psi_res}), within the latitude range of the re-entrant channel, the integral is taken over all longitudes. We define the pycnocline depths $D_n$ and $D_w$ as $\mathcal{Z}^n(0^\circ,\hat{b})$ and $\mathcal{Z}^w(0^\circ,\hat{b})$, respectively. The pycnocline depth in the MITgcm is thus interpreted as the depth of the buoyancy level separating the upper limb water masses from the lower overturning water masses.
	
	Finally, following \cite{jones2016interbasin}, the interbasin exchange flow is calculated as:
	\begin{equation}
		\psi_g(\tilde{b})=a\int_{\theta_s}^{\theta_c}\int_{-H}^0 (u^\dagger_1-u^\dagger_2) \mathcal{H}[b(\lambda,\theta,z,t)-\tilde{b}] dz d\theta,
		\label{eq:psig_MITgcm}
	\end{equation}
	where $\theta_s=70^\circ$S and $\theta_c=50^\circ$S and, $u^\dagger_1$ and $u^\dagger_2$ are the residual zonal velocities evaluated at the eastern boundary of the narrow and wide basin, respectively. Again, for comparison with the CTBM, $\psi_g$ in~(\ref{eq:psig_MITgcm}) is evaluated at $\hat{b}$.
	
	\subsection{Overturning and the hyrdological cycle}
\begin{figure}
		\captionsetup{justification=centering}
		\centering
		\includegraphics[width=\linewidth]{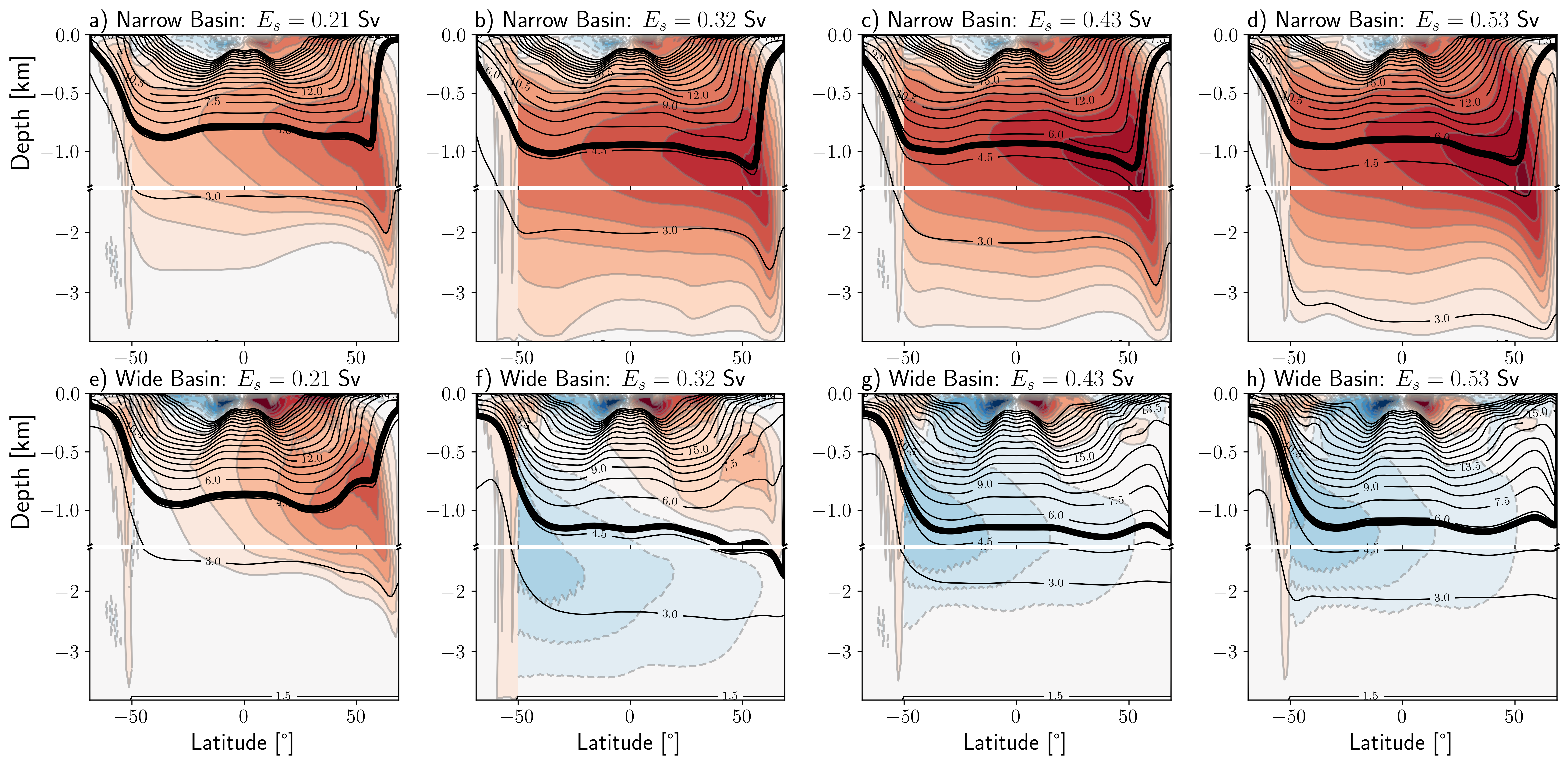}
		\caption{Residual overturning streamfunction in narrow basin ($\psi^n$; panels (a)-(d)), and wide basin ($\psi^w$; panels (e)-(f)) for four different values of $F_0$. Gray streamfunction contour intervals are spaced by $2$~Sv. Black buoyancy contours lines are spaced by $1.5\times 10^{-3}$~m~s$^{-2}$. Thick black contour indicates the buoyancy level ($\hat{b}$) used to compute the pycnocline depth and sinking strength, and equals $4.7\times 10^{-3}$, $4.7\times 10^{-3}$,  $5.2\times 10^{-3}$ and $5.7\times 10^{-3}$~m~s$^{-2}$, for $E_s=0.21, 0.32, 0.43$ and $0.53$~Sv, respectively.}
		\label{fig:MIT_Esa_vals}
	\end{figure}

	We run the model for four values of $F_0$: $0.4, 0.6, 0.8,$~and $1.0$~m year$^{-1}$. Using equation~(\ref{eq:Es_converter}), these hydrological forcing amplitudes correspond to $E_s$ values of $0.21, 0.32, 0.42,$~and $0.53$~Sv, respectively. Fig.~\ref{fig:MIT_Esa_vals} shows the resulting $\psi^n$ and $\psi^w$, mapped to the height coordinates $\mathcal{Z}^n$ and $\mathcal{Z}^w$, respectively. Table~\ref{tab:mitgcm_compar} lists the corresponding values of $\psi_n^i$, $\psi_g$, $D_n$, and $D_w$ diagnosed from the MITgcm and CTBM for each $E_s$.
	
	For weak hydrological forcing ($F_0 = 0.4$ m year$^{-1}$), the MITgcm solution exhibits interhemispheric overturning in both the narrow and wide basins. In each basin, the buoyancy surface separating the upper and lower overturning limbs outcrops in the channel and the northern part of the domain, thereby permitting pole-to-pole adiabatic flow \citep{wolfe2011adiabatic}. In the CTBM framework, this configuration corresponds to the symmetric overturning (NW) state. The sinking in the narrow basin is stronger and associated with a shallower pycnocline than in the wide basin (Table~\ref{tab:mitgcm_compar}).
	
	For $F_0 > 0.4$ m year$^{-1}$, all experiments exhibit an asymmetric overturning configuration, with sinking confined to the narrow basin, analogous to the N-state in the CTBM framework. In this regime, only the narrow basin shares outcropping isopycnals—separating the upper and lower overturning limbs—with the channel. Consequently, no adiabatic pathway exists to sustain a pole-to-pole circulation in the wide basin. 
	
	For all values of $F_0$, the separating isopycnal lies deeper in the wide basin than in the narrow basin, reflecting the larger area of the wide basin. According to~(\ref{eq:psig}), this implies an interbasin transport from the wide to the narrow basin, consistent with the diagnostics in Table~\ref{tab:mitgcm_compar}.
	
	In the asymmetric sinking regime, the CTBM and MITgcm agree well on the pycnocline depths, the strength of northern sinking, and the interbasin transport. In both basins, the pycnocline depth decreases as $E_s$ increases, while the sinking strength in the active basin strengthens with $E_s$. This behavior is consistent with the adiabatic salt–advection feedback in the quasi-adiabatic limit, in which an enhanced hydrological cycle intensity increases the range of overlapping isopycnals between the two poles \citep{wolfe2014salt}. However, consist with the CTBM, the sensitivities of the pycnocline depth and sinking strength remain modest.
	
	In the symmetric sinking regime, diagnostics from the MITgcm and CTBM are in good agreement. Both models predict substantially shallower pycnoclines in the two basins and a reduced depth difference between them, which leads to weaker interbasin geostrophic transport (Table~\ref{tab:mitgcm_compar}). These results are consistent with Fig.~\ref{fig:bif_Eib}b, where the CTBM NW-state is characterized by a minimized pycnocline depth due to active sinking in both basins. However, the CTBM overestimates the sinking strength in the wide basin—and consequently the total sinking $\psi_n^t$—by more than 2~Sv relative to the MITgcm. This discrepancy may arise from the somewhat arbitrary choice of $\hat{b}$ in the MITgcm under a double-sinking configuration. Nevertheless, the overall qualitative characteristics of this regime remain consistent with the CTBM.
	
	\begin{table}[t]
		\centering
		\caption{Model diagnostics in MITgcm and CTBM for four different values of $E_s$. }
		\label{tab:mitgcm_compar}
		\footnotesize
		\renewcommand{\arraystretch}{1.1}
		\begin{tabular}{|ll| ll ll|}
			\hline
			& $F_0$ [m year$^{-1}$] & $0.4$& $0.6$& $0.8$& $1.0$\\
			& $E_s$ [Sv]&  $0.21$ & $0.32$& $0.43$ & $0.53$ \\
			\hline
			\hline
			MITgcm & $D_n$ [m] & $781$& $936$& $925$& $897$ \\
			&$D_w$ [m]&$857$ & $1164$& $1144$ & $1100$ \\
			& $\psi_n^n$ [Sv] & $8.8$& $13$& $14.5$& $14.6$ \\
			&      $\psi_n^w$ [Sv]& $5.5$& $-2.8$& $-4.3$&  $-4.2$\\
			&      $\psi_g$ [Sv]& $3.5$& $8.8$& $9.9$& $10$ \\
			\hline
			CTBM & $D_n$ [m]& $781$& $944$ &$911$ & $884$\\
			&$D_w$ [m]& $871$&$1176$& $1153$&  $1134$\\
			& $\psi_n^n$ [Sv]& $8.3$& $14$& $14.3$&  $14.5$\\
			&      $\psi_n^w$ [Sv]& $7.9$& $0$ & $0$&  $0$~\\
			&      $\psi_g$ [Sv]& $2.7$& $8.9$ & $9.1$&  $9.2$\\
			\hline
		\end{tabular}
	\end{table}	
	
	\subsection{Multi-stability: Role of $E_s$ and $E_{ib}$}
	Fig.~\ref{fig:cont_saddle_Esa} shows that all three equilibria may coexist at $E_{ib}=0$~Sv, depending on the value $E_s$. A similar situation can occur in the MITgcm, implying that the solutions in Fig.~\ref{fig:MIT_Esa_vals} may depend on the initial conditions. If such dependence exists, at which value of $E_{ib}$ does the solution become unique? Addressing this question is essential for understanding the role of interbasin freshwater fluxes in explaining deep water formation in the Atlantic and its relative absence in the Pacific.
	
\begin{figure}
		\captionsetup{justification=centering}
		\centering
		\includegraphics[width=0.5\linewidth]{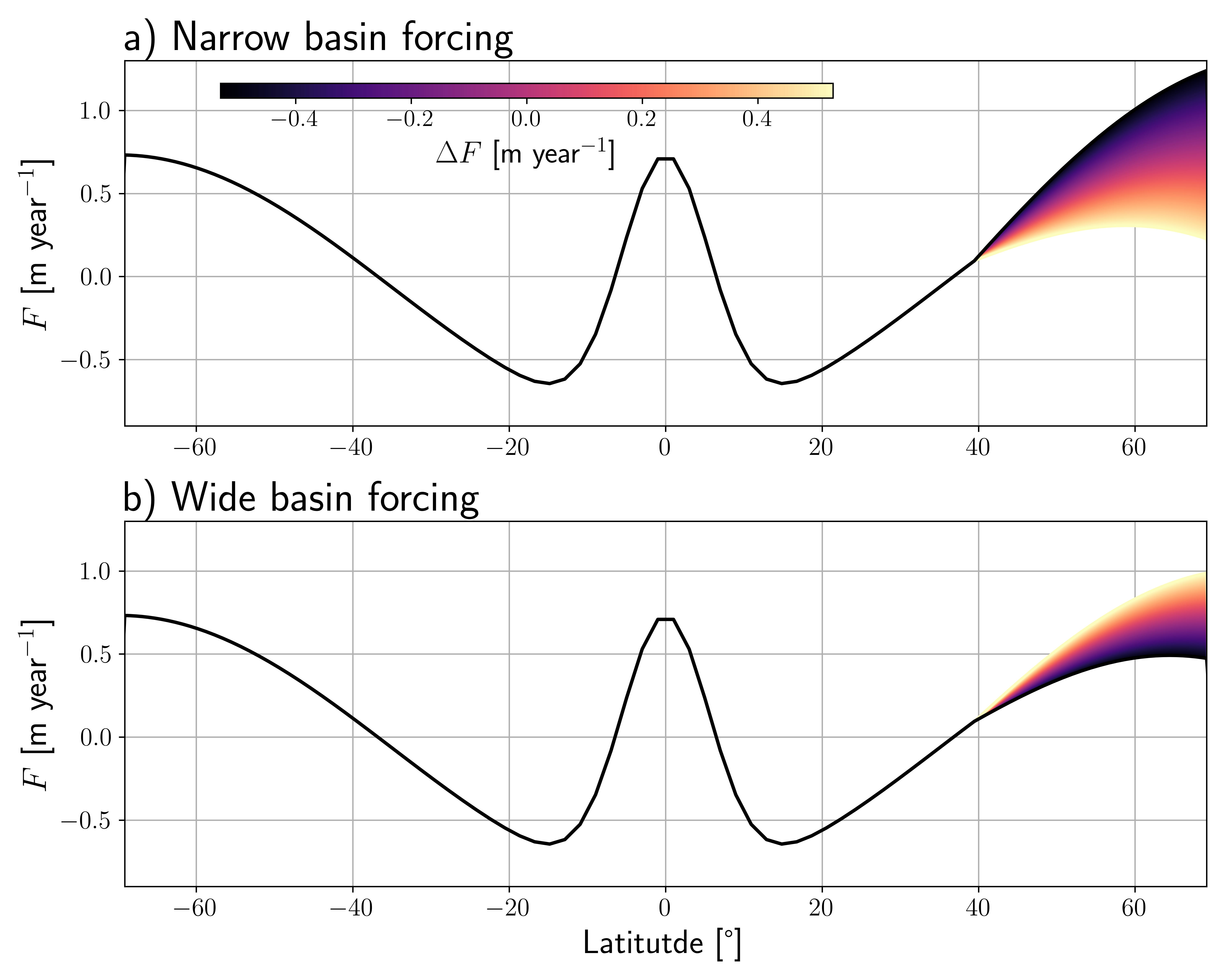}
		\caption{Freshwater flux forcing, with $F_0=0.8$~m year$^{-1}$, for various values of $\Delta F$ in the (a) narrow and (b) wide basin, with.}
		\label{fig:MIT_forc_ramp}
	\end{figure}

	To obtain a rough sketch of the stability landscape in the MITgcm, and its dependence on $E_s$, we introduce an inter-basin freshwater flux defined by including a longitudinally asymmetric surface freshwater flux in~(\ref{eq:fw_flux_MIT}), reading:
	\begin{equation}
		\begin{split}
			(P-E)_{\mathrm{asym}}(\theta,\lambda) =\\
			\begin{cases}
				0, & \theta < \theta_m,\\[1mm]
				-\Delta F \frac{\theta - \theta_m}{\theta_n - \theta_m}, & \theta \ge \theta_m, \ \lambda \in [\lambda_1^n, \lambda_2^n],\\[1mm]
				\gamma \Delta F \frac{\theta - \theta_m}{\theta_n - \theta_m}, & \theta \ge \theta_m, \ \lambda \in [\lambda_1^w, \lambda_2^w],
			\end{cases}
		\end{split}
		\label{eq:fw_asym}
	\end{equation}
	where $\Delta F$ is the amplitude of the freshwater flux asymmetry, $\theta_m = 40^\circ$N, and $\gamma = 1/2$ accounts for the ratio of the surface area of the narrow basin to that of the wide basin, ensuring that equation~(\ref{eq:fw_asym}) integrates to zero over all longitudes. For $\Delta F > 0$, there is a net moisture transport from the narrow to the wide basin, and vice versa for $\Delta F < 0$. Adding equation~(\ref{eq:fw_asym}) to equation~(\ref{eq:fw_flux_MIT}) yields the total freshwater forcing, including both the zonally symmetric and asymmetric components. Figure~\ref{fig:MIT_forc_ramp} shows this forcing in the narrow and wide basins for different values of $\Delta F$.
	
	The surface integral of~(\ref{eq:fw_asym}) over the narrow basin corresponds to the value of $E_{ib}$. For each value of $E_s$ listed in Table~\ref{tab:mitgcm_compar}, the model is initialized with $\Delta F = 0.52$ and $-0.52$~m~year$^{-1}$, corresponding to $E_{ib} = 0.1$ and $-0.1$~Sv, respectively. These freshwater forcing profiles guarantee asymmetric sinking states with sinking localized in the narrow basin and wide basin, respectively. Once a steady state is reached, $E_{ib}$ is linearly ramped from $0.1$ to $-0.1$~Sv and from $-0.1$ to $0.1$~Sv over $10^4$ model years, defining the backward and forward hosing experiments. The slow ramping ensures quasi-equilibrium conditions throughout the experiments.
	
	The $E_{ib}$ values of the saddle points $L_n^-$ and $L_w^+$ in the MITgcm simulations are estimated by applying a Savitzky–Golay filter, with a $1\times 10^{-2}$~Sv ($=500$~year) window size, to $\psi_n^n$ and $\psi_n^w$ along the forward and backward branches, respectively. The saddle is identified as the value of $E_{ib}$ at which the magnitude of the derivative of the filtered signal first exceeds $100$~Sv~Sv$^{-1}$ ($=2\times 10^{-3}$~Sv~year$^{-1}$). Although this slope threshold is arbitrary, it provides a useful first-order estimate that can be consistently compared across the different simulations.
	
\begin{figure}
		\captionsetup{justification=centering}
		\centering
		\includegraphics[width=\linewidth]{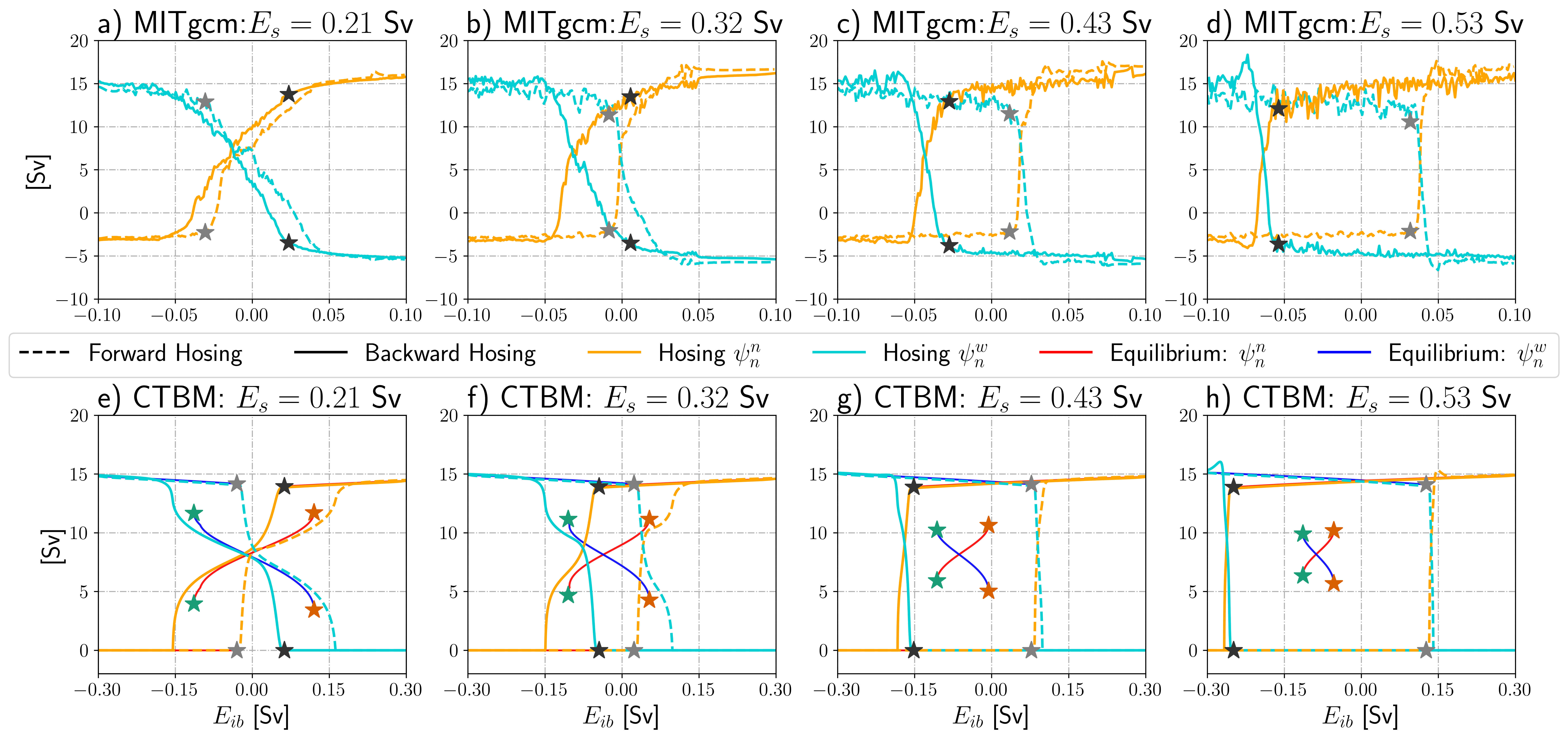}
		\caption{Backward (thick lines) and forward (dashed lines) hosing experiments performed with the MITgcm (a–d) and CTBM (e–h) for four values of $E_s$. To facilitate interpretation, the northern sinking strength in the MITgcm is computed at a fixed buoyancy level, $\hat{b} = 5.2 \times 10^{-3}$ m s$^{-2}$. For reference, the bifurcation diagrams (i.e., equilibrium solutions) of the CTBM and their corresponding saddle-node bifurcations for the same values of $E_s$ are shown in the lower panels. For these bifurcation diagrams, the same legend as in Fig.~\ref{fig:bif_Eib}a applies. Estimates of $E_{ib}(L_n^-)$ and $E_{ib}(L_w^+)$ in the MITgcm are indicated by stars.}
		\label{fig:MIT_Esa_hos}
	\end{figure}

	Fig.~\ref{fig:MIT_Esa_hos}a-d shows the results of the hosing simulations for four values of $E_s$ in the MITgcm. For all values of $E_s$, the forward and backward branches are initialized in the W- and N-states, respectively. These asymmetric states exhibit similar sinking strengths, as the total overturning remains approximately invariant with respect to the location of sinking \citep{jones2016interbasin}. Starting from these asymmetric states, increasing $E_{ib}$ weakens sinking in the wide basin along the forward branch, while decreasing $E_{ib}$ weakens sinking in the narrow basin along the backward branch. 
	
	For $E_s = 0.21$~Sv, both the backward and forward branches converge to the NW-state for sufficiently small values of $E_{ib}$, i.e. beyond the approximate saddle points at $E_{ib}(L_n^-) \approx 2.4 \times 10^{-2}$~Sv and $E_{ib}(L_w^+) \approx -0.03 \times 10^{-2}$~Sv, respectively (Fig.~\ref{fig:MIT_Esa_hos}a). Indeed, Fig.~\ref{fig:MIT_Esa_vals}a,e confirms that the NW-state is stable at $E_{ib} = 0$. For this value of $E_s$, the MITgcm exhibits hysteresis behavior arising from the coexistence of asymmetric and symmetric sinking states. Specifically, two regions of bistability can be identified: one where the N- and NW-states coexist, and another where the W- and NW-states coexist. However, because the transitions occur between asymmetric and symmetric configurations, differences in sinking strength between the forward and backward branches—and the resulting hysteresis—remain subtle within each basin.
	
\begin{figure}[t!]
		\centering
		\includegraphics[width=0.5\linewidth]{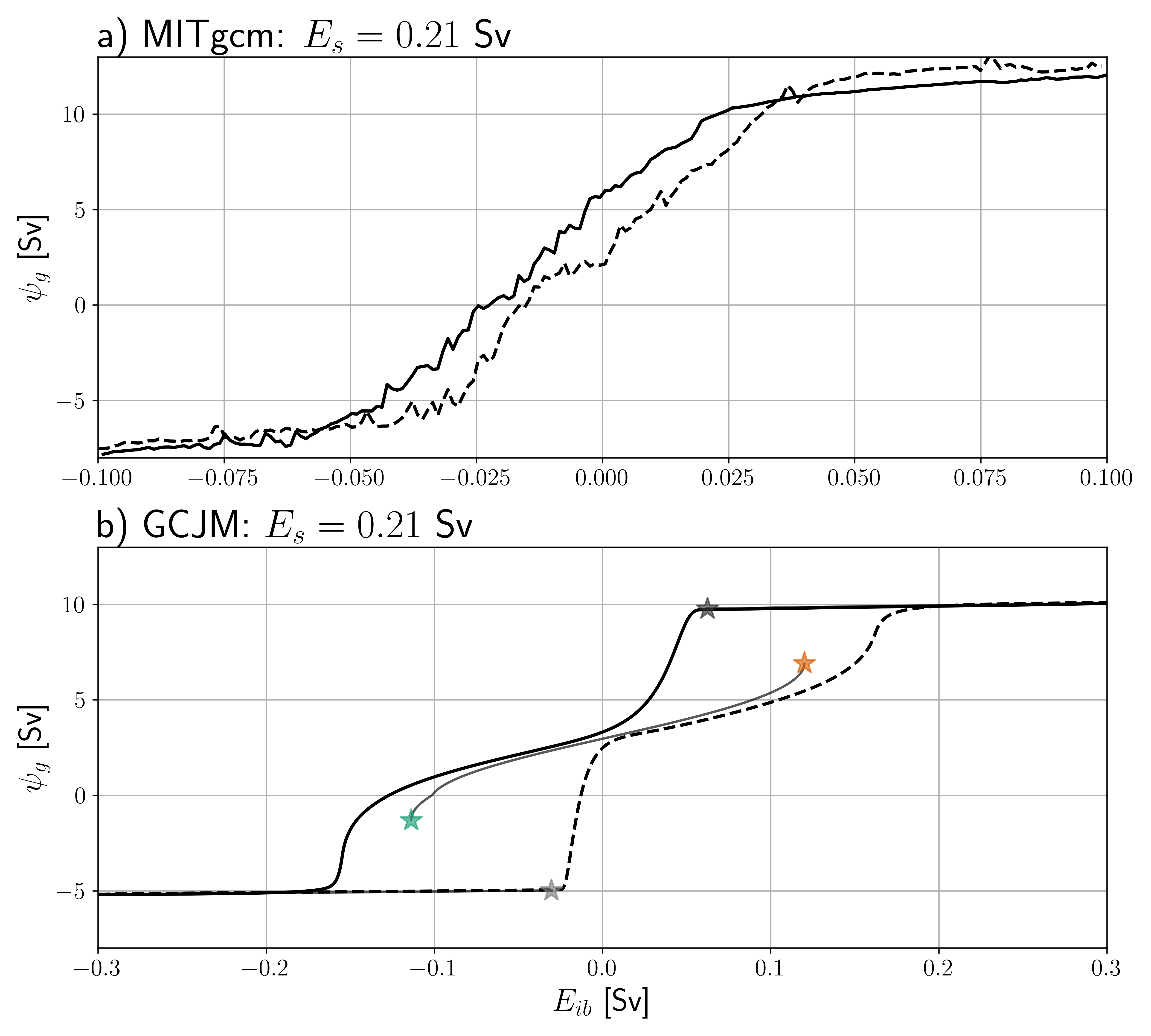}
		\caption{Interbasin geostrophic transport in backward (thick lines) and forward (dashed lines) hosing experiment for $E_s=0.21$~Sv. In the MITgcm, $\psi_g$ is evaluated at $\hat{b}=5.2\times 10^{-3}$ m s$^{-2}$}
		\label{fig:MIT_psig}
	\end{figure}

	Within the NW-state, the sinking strength in both basins, which behave like a seesaw, shows a stronger dependence on $E_{ib}$ compared to the asymmetric sinking state (Fig.~\ref{fig:MIT_Esa_hos}a,e). As outlined in Section~\ref{S:MM}\ref{S:RefSol_GCJM}, this arises because, in the NW-state, $E_{ib}$ forces changes in sinking strength of both basins, resulting in a strong response of $\psi_g$ to changes in $E_{ib}$. Fig.~\ref{fig:MIT_psig}a confirms that the strength of $\psi_g$ increases considerably with $E_{ib}$ in the NW-state of the MITgcm, while remaining only weakly dependent on $E_{ib}$ in the asymmetric sinking states. 
	
	For $E_s = 0.32$~Sv, the stability range in $E_{ib}$ of the asymmetric N- and W-states increases along the backward and forward branches, respectively (Fig.~\ref{fig:MIT_Esa_hos}b). Specifically, the approximated saddle points are now located found at $E_{ib}(L_n^-)\approx 5\times 10^{-3}$~Sv and $E_{ib}(L_w^+)\approx -8\times10^{-3}$~Sv. Moreover, the transition between the two asymmetric sinking states occurs over a narrower range of $E_{ib}$, indicating a contraction of the stability interval of the NW-state relative to the $E_s = 0.21$~Sv hosing experiment (Fig.~\ref{fig:MIT_Esa_hos}a). 
	
	In addition, the transition from the N-state to the W-state along the backward branch spans a larger range of $E_{ib}$ values than the reverse transition along the forward branch. In the MITgcm, the N-to-W transition occurs between $E_{ib}(L_n^-) \approx 5 \times 10^{-3}$ and $-5 \times 10^{-2}$~Sv, whereas the W-to-N transition spans $E_{ib}(L_w^+) \approx -8 \times 10^{-3}$ to $4 \times 10^{-2}$~Sv. Consequently, the transition interval for the N-to-W transition is wider by approximately $7 \times 10^{-3}$~Sv. This asymmetry suggests that the NW-state has shifted toward negative $E_{ib}$ as $E_s$ increases.
		
	With increasing $E_s$, the asymmetric N- and W-states occupy a larger stability range in $E_{ib}$ along the backward and forward branches, respectively (Fig.~\ref{fig:MIT_Esa_hos}c,d). In addition, the transitions between the two asymmetric overturning states become sharper in $E_{ib}$-space as $E_s$ increases. This behavior is consistent with a continued contraction of the NW-state with increasing $E_s$, resulting in a weaker attraction toward this state during transitions between the asymmetric regimes. However, the slower transition from the N-state to the W-state, compared to the reverse transition, remains evident (Fig.~\ref{fig:MIT_Esa_hos}c) and again reflects the shift of the NW-state toward more negative $E_{ib}$ values. This discrepancy becomes less pronounced for $E_s=0.53$~Sv, suggesting that the NW-state becomes very narrow and is located far from the saddle-node bifurcations $L_n^-$ and $L_w^+$.
	
	Fig.~\ref{fig:MIT_Esa_hos}e-h shows the result of similar hosing experiments performed in the CTBM. A first important discrepancy between the two models is that the stability landscape of the MITgcm spans a smaller range of $E_{ib}$ values than in the CTBM, suggesting a greater sensitivity to interbasin freshwater fluxes. This enhanced sensitivity arises from the formulation in equation~(\ref{eq:fw_asym}), in which the freshwater forcing is disproportionately applied in the polar convective regions (Fig.~\ref{fig:MIT_forc_ramp}). As a result, deep water formation in the passive basin—and thus a regime shift—is triggered more readily in the MITgcm. A more spatially uniform distribution of the freshwater forcing in the MITgcm would likely reduce this sensitivity and improve agreement with the CTBM.
	
	Nevertheless, the GOC stability landscape of the MITgcm is broadly consistent with the CTBM. For low $E_s$, the system exhibits a monostable symmetric overturning state, requiring relatively large magnitudes of $E_{ib}$ to break this symmetry and localize sinking in a single basin (Fig.~\ref{fig:MIT_Esa_hos}e). As $E_s$ increases, the asymmetric states become stable over a broader range of $E_{ib}$, while the stability range of the NW-state shrinks and shifts toward negative $E_{ib}$. As a result, the transition from the N-state to the W-state becomes more gradual than the reverse transition (Fig.~\ref{fig:MIT_Esa_hos}f,g). In this moderate $E_s$ regime, sinking preferentially localizes in the most evaporative basin.  For high $E_s$ (Fig.~\ref{fig:MIT_Esa_hos}h), the stability ranges of both asymmetric states expand further, requiring very large magnitudes of $E_{ib}$ before sinking becomes uniquely localized in one basin. Consequently, the notion that the most evaporative basin always hosts sinking no longer holds in this regime. Moreover, the NW-state has shrunk to the extent that it can no longer influence the transitions between asymmetric overturning states, explaining the sharp transitions at large $E_s$. 
	
\begin{figure}
		\captionsetup{justification=centering}
		\centering
		\includegraphics[width=0.5\linewidth]{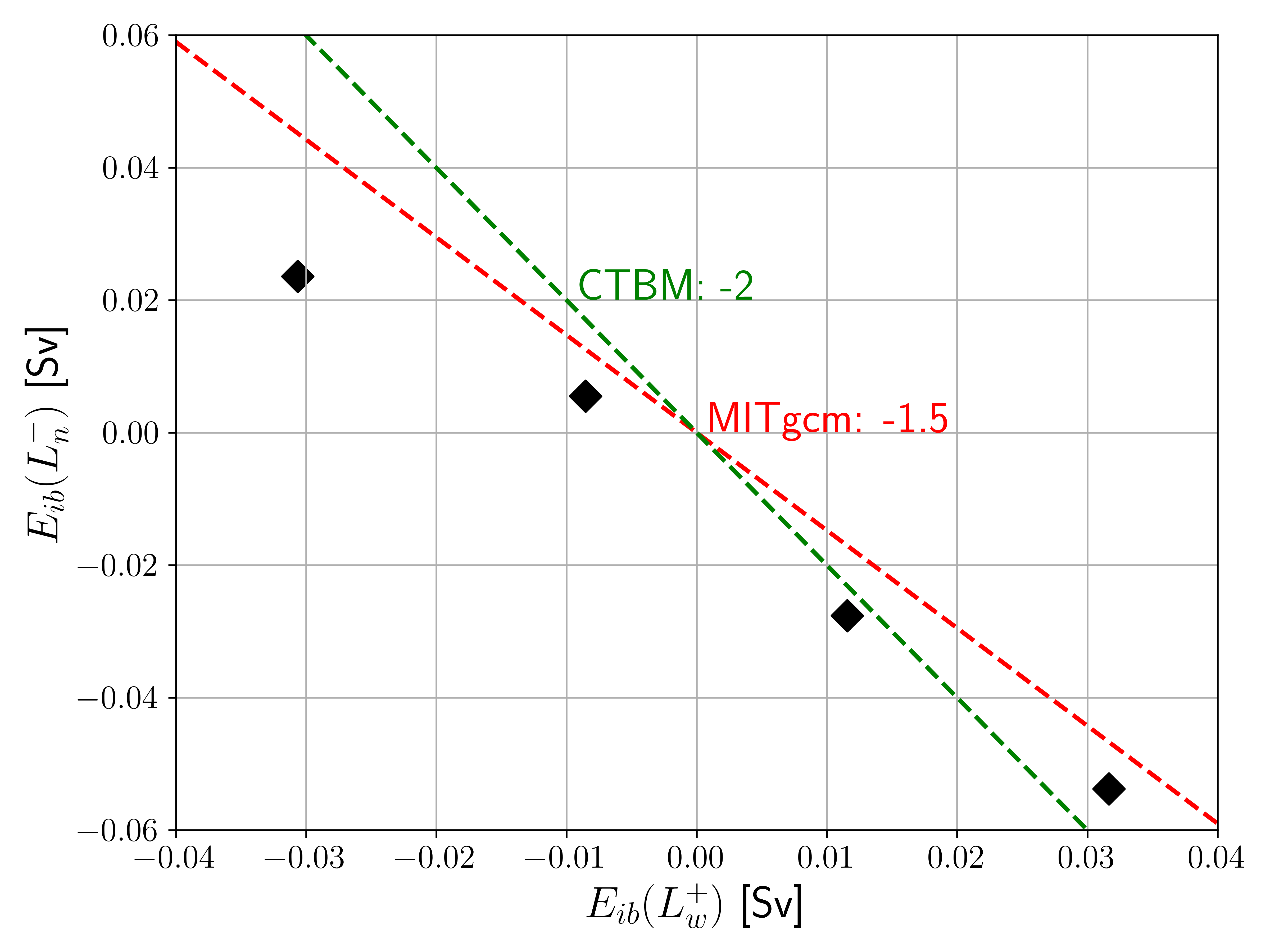}
		\caption{$E_{ib}$ values at saddle points of asymmetric sinking states in MITgcm. Dashed shows reference slope lines for the CTBM (green) and MITgcm (red). The evaluation of the slope in the MITgcm slightly depends on method used to estimate $E_{ib}(L_n^-)$ and $E_{ib}(L_w^+)$. However, the reduced dependence relative to the CTBM is consistent across the various methods applied.}
		\label{fig:MIT_Ln_Lp}
	\end{figure}

	The physical mechanisms underlying the shift of these bifurcation points in the CTBM were explored in Section~\ref{S:M}\ref{S:Es_Sens} and, given the good qualitative agreement, likely also apply to the MITgcm. One clear consequence of the basin-size asymmetry is the more rapid expansion of the stable $E_{ib}$ range in the N-state compared to the W-state. In the CTBM, we showed that $E_{ib}(L_n^-) = -2E_{ib}(L_w^+)$. For the MITgcm, we find $E_{ib}(L_n^-) \approx -1.5E_{ib}(L_w^+)$ (Fig.~\ref{fig:MIT_Ln_Lp}). In the CTBM, this scaling arises because the wide basin is twice as wide as the narrow basin, with gyre mixing strength scaling proportionally with basin width. As a result, the salinity of the northern box in the narrow basin (W-state) is twice as sensitive to changes in $E_{ib}$ as that in the wide basin (N-state) (Fig.~\ref{fig:bif_Eib}c). The smaller slope magnitude in the MITgcm indicates that this sensitivity contrast between basins is reduced relative to the CTBM. One possible explanation is that stronger northward subtropical gyre velocities in the wide basin \citep{jones2017size} enhance gyre-induced salt transport into high latitudes relative to the narrow basin. In the absence of wide-basin sinking, this enhanced salt transport promotes the onset of a PMOC (i.e., $\rho_n^w - \rho_{t_s}^w > 0$) at larger (more positive) freshwater flux asymmetries. Because the box model assumes identical northern gyre mixing strength per unit basin width, this effect is not represented but could be incorporated by setting $r_n^w > 2r_n$. This is left for future work.
	
	\section{Summary \& Discussion}\label{S:S&D}
	Many studies have linked the absence of NPDW formation, and the associated PMOC, to the higher atmospheric freshwater flux over the Pacific compared to the Atlantic, where NADW formation sustains an active AMOC \citep[e.g.,][]{menviel2012removing,su2018difference}. However, the potential multistability of the GOC complicates a mechanistic understanding of how freshwater flux asymmetry biases sinking toward a particular basin, and how this preference depends the strength of the hydrological cycle. Here, building on earlier work \citep{cimatoribus2014meridional,jones2016interbasin}, we develop a conceptual model of the global ocean as a two-basin system—one wide and one narrow—connected by a zonally periodic re-entrant channel.
	
	We identify three dynamically distinct equilibria as a function of the interbasin surface freshwater flux ($E_{ib}$): (i) sinking confined to the narrow basin, (ii) sinking confined to the wide basin (together referred to as asymmetric sinking states), and (iii) sinking occurring in both basins (symmetric sinking state). All three states have been identified in earlier studies using more comprehensive climate models \citep[e.g.,][]{hu2021influence}, and are reproduced here with the MITgcm. We do not find a state in which sinking is absent in both basins. This absence follows directly from the choice of control parameter, which imposes opposing surface salinity anomalies in the two basins.
	
	When one basin is considerably more evaporative than the other, it generally hosts sinking. The symmetric overturning state remains stable only under small freshwater flux asymmetries. However, the precise meaning of “considerably” and “small” depends strongly on the strength of the hydrological cycle, which in this study is interpreted as the longitudinally symmetric component of the surface freshwater flux. As the hydrological cycle intensifies, the asymmetric sinking states remain stable over an increasingly wide range of flux asymmetries, whereas the symmetric overturning state persists over a progressively narrower range, eventually disappearing altogether. These findings, consistent with experiments in the MITgcm, have important implications for the influence of freshwater flux asymmetries across different climate regimes.
	
\begin{figure}
		\captionsetup{justification=centering}
		\centering
		\includegraphics[width=0.5\linewidth]{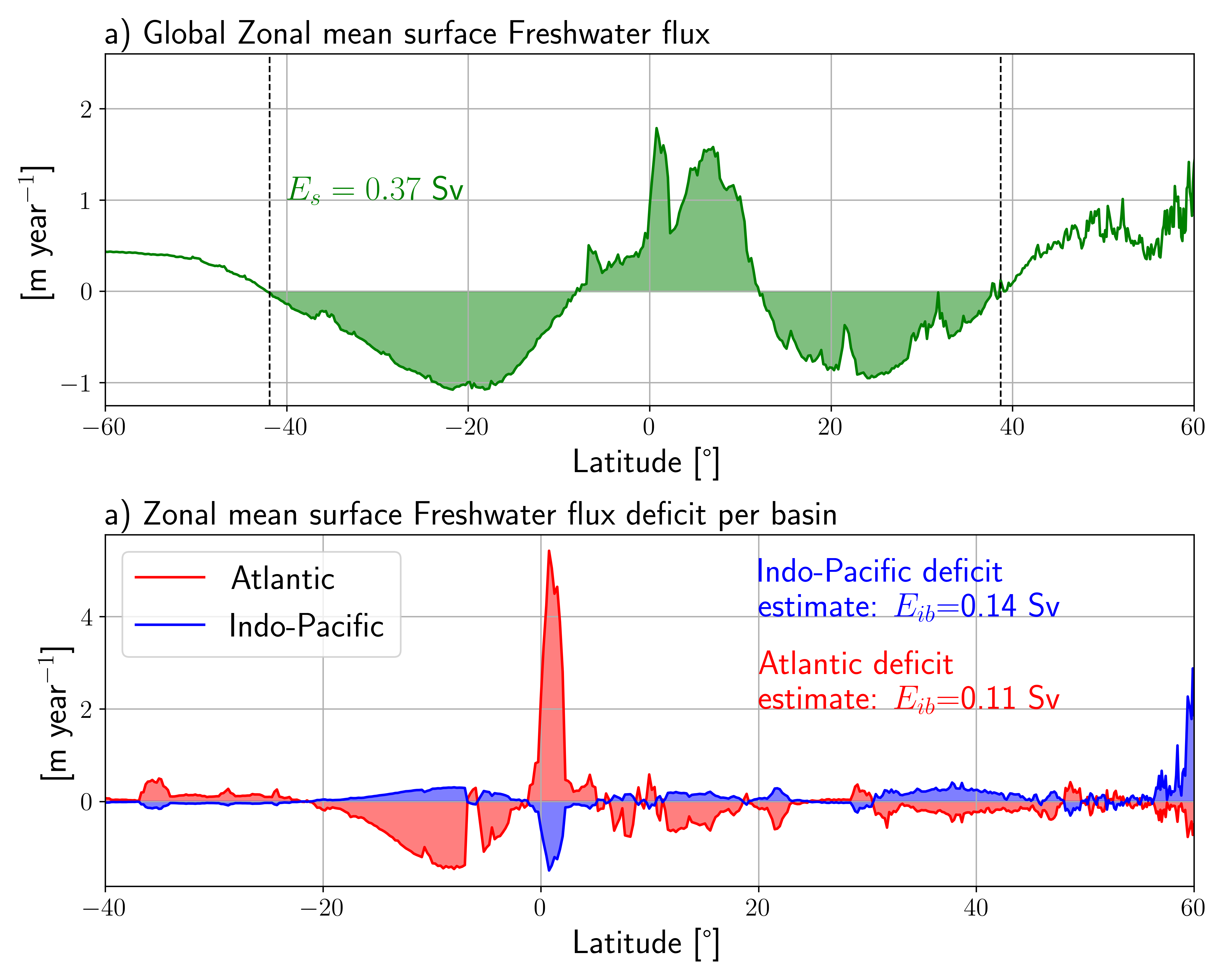}
		\caption{Estimates of (a) $E_s$ and (b) $E_{ib}$ are based on the time-averaged (1991–2018) surface freshwater flux from the ORAS5 reanalysis dataset, which includes precipitation, evaporation, runoff and sea-ice fluxes (\url{https://os.copernicus.org/articles/15/779/2019/}). The value of $E_s$ is obtained by averaging the global zonal-mean surface freshwater flux between the two dashed vertical lines in panel (a) and multiplying by the total area of the Atlantic Ocean within this latitude range. The interbasin freshwater flux $E_{ib}$ is estimated by calculating the difference between the zonal mean in each basin and the global zonal mean, followed by integration over the full basin. Panel (b) shows that these two deficits are strongly inversely correlated, indicating a net longitudinal moisture transport from the Indo-Pacific to the Atlantic. The $E_{ib}$ values for the two basins differ slightly, reflecting that part of the excess moisture may be exported to other seas or oceans.}
		\label{fig:fw_flux_ORAS5}
	\end{figure}

	In the present-day climate, Fig.~\ref{fig:fw_flux_ORAS5} indicates a hydrological cycle intensity of roughly $0.37$~Sv and an interbasin freshwater flux of approximately $E_{ib}\approx 0.1-0.15$~Sv, implying that the Atlantic is more evaporative. In the conceptual model stability landscape (Fig.~\ref{fig:cont_saddle_Esa}a), these values place the climate in the moderate-intensity hydrological cycle regime. In this regime, asymmetric sinking states become stable even at relatively low magnitudes of interbasin freshwater flux, with sinking preferentially occurring in the most evaporative basin. Both the conceptual model and the MITgcm suggest that $E_{ib} \approx 0.1-0.15$~Sv would confine sinking to the Atlantic, consistent with present-day observations. However, the GOC remains highly sensitive to the sign and magnitude of $E_{ib}$, in agreement with studies showing strong responses to changes in interbasin freshwater flux asymmetries—for example, due to variations in land orography \citep{schmittner2011effects,sinha2012mountain}. Other work suggests that a PMOC may have developed during the last glacial termination when meltwater discharges in the Atlantic effectively reduced the interbasin freshwater flux asymmetry \citep{okazaki2010deepwater}. Overall, our work, with the support of earlier studies, suggests the existence of the present-day AMOC can largely be attributed to the Atlantic being more evaporative than the Pacific.
	
	In climate states with a weak hydrological cycle, the symmetric overturning state is the only stable equilibrium, even under relatively large interbasin freshwater flux asymmetries. This occurs because high-latitude surface freshwater forcing is too weak to prevent overlapping outcropping isopycnals in both basins. Consequently, interbasin freshwater flux asymmetry has little influence on the sinking location. A weakened hydrological cycle has been proposed for the warm mid-Pliocene climate \citep{burls2017active}, during which proxy records suggest that the GOC may have maintained a symmetric overturning state \citep{burls2017active,ford2022sustained}. However, the interpretation of these proxies remains debated \citep{novak2024isotopic}, with multi-model comparisons indicating a potentially strengthened hydrological cycle during the mid-Pliocene \citep{han2021evaluating}.
	
	Finally, under a very high-intensity hydrological cycle, the symmetric sinking state is stable over only a very narrow—or even vanishing—range of interbasin freshwater flux values, whereas the asymmetric states remain stable across an increasingly wide range of surface freshwater flux asymmetries. Notably, these asymmetric states persist even when the passive basin is considerably more evaporative than the active basin. This implies that both AMOC and PMOC configurations are possible, with the freshwater flux asymmetry exerting only a weak control on the sinking location unless its magnitude is very large. Consequently, in this regime, the realized state will generally depend on the initial conditions.
	
	Assuming an approximate increase in $E_s$ of $7\%$ per degree of global warming \citep{held2006robust}, a $4^\circ$C warming yields $E_s \approx 0.48$~Sv. In the stability diagram (Fig.~\ref{fig:cont_saddle_Esa}a), this corresponds to a regime in which AMOC and PMOC states coexist if $E_{ib}$ remains unchanged—an assumption that may be difficult to justify. Hence, under extreme global warming, our conceptual model suggests that a PMOC may be spontaneously initiated and remain stable thereafter. Such spontaneous transitions have been identified in extreme emission scenarios \citep{curtis2024spontaneous}.
	
	Alongside the dependencies on hydrological cycle intensity and interbasin freshwater flux, the stability landscape of the two-basin system is asymmetrically organized around a fully symmetric surface forcing, reflecting the difference in basin size. In particular, for moderate to high hydrological cycle intensities, the AMOC is generally uniquely realized—that is, independent of initial conditions—at lower magnitudes $E_{ib}$ than the PMOC. This preference arises because the larger size of the Pacific renders its sea surface salinity less sensitive to perturbations in interbasin freshwater flux. Moreover, under these conditions, the symmetric overturning state preferentially occurs when the Pacific is more evaporative than the Atlantic.  This reflects the dynamics of the symmetric overturning state, in which interbasin transport is preferentially directed from the Atlantic to the Pacific. This increases Atlantic salinity and allows the Atlantic salinity-advection feedback to dominate at lower magnitudes of freshwater flux asymmetry than in the Pacific.
	
	Our results rely on several simplifying assumptions regarding oceanic and atmospheric circulation. First, atmospheric and sea-ice freshwater fluxes depend on the ocean state—a coupling that has been shown to play an important role in GOC transitions \citep[e.g.,][]{su2018difference}—but is neglected in this study. Second, we consider only two ocean basins with closed gateways. The presence of ocean gateways has been shown to influence AMOC–PMOC seesaw behavior \citep{hu2012pacific} and the sensitivity of the GOC to freshwater perturbations \citep{hu2012role,hu2021influence}. These processes could, in principle, be incorporated in an extended version of the model \citep[e.g.,][]{soons2026effect}. Third, our model configuration assumes that the two continents bounding the longitudinal extent of the semi-enclosed basins extend equally far southward. In reality, however, the eastern boundary of the Atlantic extends only to approximately $34^\circ$S, whereas South America reaches about $55^\circ$S. This asymmetry has been argued to strengthen the salt-advection feedback in the Atlantic relative to the Pacific \citep{vanderborght2025feedback}, thereby enhancing Atlantic salinification and favoring the existence of an AMOC over a PMOC \citep{nilsson2013ocean,cessi2017warm}.
	
	However, these simplifications offer the advantage of a transparent isolation of the role of atmospheric freshwater fluxes in shaping the ocean state. Nevertheless, this study only scratches the surface of the rich behavior and dependencies that the conceptual model may exhibit. The roles of gyres, eddies, interhemispheric temperature contrasts, and Southern Ocean surface zonal wind stress can be investigated in a similar framework and are expected to play important roles in determining the preference for an AMOC over a PMOC, as well as how this preference varies across different climate states.
	\section*{Acknowledgments} 
	E.V.,  O.M.  and  H.A.D. are funded by the European Research Council through the ERC-AdG project  TAOC (PI: Dijkstra, project 101055096). The MITgcm simulations and the analysis of all the model output was conducted on the Dutch National Supercomputer Snellius within NWO-SURF project 2024.013.  

	\section*{Availability Statement}
	The model code and analysis scripts will be made available on Zenodo upon
	publication.
	\section*{Appendix A: Salinity Equations}
	The salinity equations of the salt content in the northern, thermocline, $t_s$, deep and southern ocean boxes, respectively read:
	\begin{align*}
		V_n^n\frac{d S_n^n}{d t}&=\psi_n^n(S_t^n-S_n^n)-(E_s^n-E_{ib})S_0+r_n^n(S_t^n-S_n^n),\\
		V_n^w\frac{d S_n^w}{d t}&=\psi_n^w(S_t^w-S_n^w)-(E_s^w+E_{ib})S_0+r_n^w(S_t^w-S_n^w),\\
		\frac{d(V_t^n S_t^n)}{dt}&=\psi_u^n S_d^n-\psi_n^nS_t^n+2E_s^nS_0+r_s^n(S_{t_s}^n-S_t^n)\\&+r_n^n(S_n^n-S_t^n)+
		\begin{cases}
			\psi_s^nS_{t_s}^n\\
			\psi_s^nS_{t}^n
		\end{cases}
		+
		\begin{cases}
			\psi_gS_{t}^w\\
			\psi_gS_{t}^n
		\end{cases},\\
		\frac{d(V_t^w S_t^w)}{dt}&=\psi_u^w S_d^w-\psi_n^wS_t^w+2E_s^wS_0+r_s^w(S_{t_s}^w-S_t^w)\\&+r_n^w(S_n^w-S_t^w)+
		\begin{cases}
			\psi_s^wS_{t_s}^w\\
			\psi_s^wS_{t}^w
		\end{cases}
		-
		\begin{cases}
			\psi_gS_{t}^w\\
			\psi_gS_{t}^n
		\end{cases},\\
		\frac{d(V_{t_s}^n S_{t_s}^n)}{dt}&=\psi_{ek}^nS_s-\psi_eS_{t_s}^n+r_s^n(S_t^n-S_{t_s}^n)-		\begin{cases}
			\psi_s^nS_{t_s}^n\\
			\psi_s^nS_{t}^n
		\end{cases},\\
		\frac{d(V_{t_s}^w S_{t_s}^w)}{dt}&=\psi_{ek}^wS_s-\psi_eS_{t_s}^w+r_s^w(S_t^w-S_{t_s}^w)-	\begin{cases}
			\psi_s^wS_{t_s}^w\\
			\psi_s^wS_{t}^w
		\end{cases},\\
		V_s\frac{dS_s}{dt}&=-(\psi_{ek}^n+\psi_{ek}^n)S_s+\psi_e^nS_{t_s}^n+\psi_e^wS_{t_s}^w\\&-(E_s^n+E_s^w)S_0+
		\begin{cases}
			\psi_s^nS_d^n\\
			\psi_s^nS_s
		\end{cases}+
		\begin{cases}
			\psi_s^wS_d^w\\
			\psi_s^wS_s
		\end{cases},\\
		\frac{d(V_d^nS_d^n)}{dt}&=\psi_n^nS_n^n-\psi_u^nS_d^n-\begin{cases}
			\psi_gS_d^n\\
			\psi_gS_d^w
		\end{cases}-\begin{cases}
			\psi_sS_d^n\\
			\psi_sS_s
		\end{cases}.
	\end{align*}
	Here the upper (lower) cases apply when the volume flux is greater (smaller) than zero. All the variables are explained in the text.
	
	\section*{Appendix B: Derivation of Equation~(\ref{eq:saddle_dependence})}
	\renewcommand{\theequation}{A.\arabic{equation}}
	\renewcommand{\thefigure}{A.\arabic{figure}}
	\setcounter{equation}{0}
	\setcounter{figure}{0}
	We derive an expression for $E_{ib}(L_n^-)$ in the adiabatic limit ($\psi_u^i=0$) from the following steady-state equations:
	\begin{subequations}
		\begin{align}
			\alpha\Delta T=\beta(S_{t_s}^w-S_{n}^w),\label{zero_dens_contrast}\\
			r_n^w(S_t^w-S_n^w)=(E_{ib}(L_n^-)+E_s^w)S_0,\label{Sn_bal_passive}\\
			(r_s^w+\psi_g)(S_{t_s}^w-S_t^w)+r_n^w(S_n^w-S_t^w)=-2E_s^wS_0,\label{St_bal_passive}\\
			\psi_s^w-\psi_g=0,\label{pyc_bal_passive}
		\end{align}
	\end{subequations}
	where we defined $\Delta T \equiv T_{t_s} - T_n$, so that equation~(\ref{zero_dens_contrast}) enforces $\rho_n^w = \rho_{t_s}^w$ for $E_{ib} = E_{ib}(L_n^-)$ (Fig.~\ref{fig:bif_Eib}d), equation~(\ref{Sn_bal_passive}) represents the salt budget of the $n^w$-box, equation~(\ref{St_bal_passive}) the salt budget of the $t^w$-box, and equation~(\ref{pyc_bal_passive}) the steady-state pycnocline volume budget in the wide basin.
	
	Rewriting equation~(\ref{St_bal_passive}) for $S_t^w$, we find:
	\begin{equation*}
		S_t^w=\frac{1}{\psi_g+r_s^w+r_n^w}\left[(\psi_g+r_s^w)S_{t_s}^w+r_n^wS_n^w+2S_0E_s^w\right]
	\end{equation*}
	which inserted in equation~(\ref{Sn_bal_passive}) yields:
	\begin{equation}
		r_n^w\left[S_{t_s}^w-S_{n}^w+\frac{1}{\psi_g+r_s^w+r_n^w}\left(r_n^w(S_n^w-S_{t_s}^w)+2S_0E_s^w\right)\right]=(E_{ib}(L_n^-)+E_s^w)S_0.
		\label{simplified_relation_Ln}
	\end{equation}
	Substituting equation~(\ref{zero_dens_contrast}) into (\ref{simplified_relation_Ln}) to replace $S_{t_s}^w - S_n^w = \alpha\Delta T/\beta$ then directly yields equation~(\ref{Lnmin_adiabatic}). An entirely analogous derivation for $E_{ib}(L_w^+)$ yields equation~(\ref{Lwmax_adiabatic}), and is omitted for brevity.
	\bibliographystyle{ametsocV6}
	\bibliography{bibliography}
\end{document}